\documentclass{article}
\usepackage{amssymb,latexsym, setspace, epsfig}
\usepackage[leqno]{amsmath}

\doublespace

\newtheorem{theorem}{Theorem}[section]
\newtheorem{lemma}[theorem]{Lemma}
\newtheorem{proposition}[theorem]{Proposition}

\newtheorem{definition}[theorem]{Definition}

\newtheorem{remark}[theorem]{Remark}
\newtheorem{assumption}[theorem]{Assumption}

\DeclareMathOperator{\esssup}{ess\, sup}
\def\define{\stackrel{\Delta}{=}}

\def\R{\mathbb{R}}
\def\I{\mathbb{I}}
\def\endproof{\hfill\diamond}

\def\sF{{\cal F}}

\def\zbar{\bar{z}}

\begin{document}

\centerline{\LARGE\bf Optimal Dynamic Portfolio with Mean-CVaR Criterion}

\hfil\break {\bf Jing Li\footnote{The findings and conclusions expressed are solely those of the author and do
not represent views of the Federal Reserve Bank of New York, or the staff of the Federal Reserve System.}},
Federal Reserve Bank of New York, New York, NY 10045, USA. Email: jing.li@ny.frb.org

\noindent {\bf Mingxin Xu}, University of North Carolina at
Charlotte, Department of Mathematics and Statistics, Charlotte, NC
28223, USA. Email: mxu2@uncc.edu


\begin{abstract} 
Value-at-Risk (VaR) and Conditional Value-at-Risk (CVaR) are popular risk measures from academic, industrial and regulatory perspectives.  The problem of minimizing CVaR is theoretically known to be of Neyman-Pearson type binary solution.  We add a constraint on expected return to investigate the Mean-CVaR portfolio selection problem in a dynamic setting: the investor is faced with a Markowitz \cite{Markowitz} type of risk reward problem at final horizon where variance as a measure of risk is replaced by CVaR.  Based on the complete market assumption, we give an analytical solution in general.  The novelty of our solution is that it is no longer Neyman-Pearson type where the final optimal portfolio takes only two values.  Instead, in the case where the portfolio value is required to be bounded from above, the optimal solution takes three values; while in the case where there is no upper bound, the optimal investment portfolio does not exist, though a three-level portfolio still provides a sub-optimal solution.
\end{abstract}

\hfill\break \textbf{Keywords:} Conditional Value-at-Risk,  
Mean-CVaR Portfolio Optimization, Risk Minimization, Neyman-Pearson Problem

\section{Introduction}\label{Section: Introduction}

The portfolio selection problem published by Markowitz \cite{Markowitz}  in 1952 is formulated as an optimization problem in a one-period static setting with
the objective of maximizing expected return, subject to the constraint of variance being bounded from above.
In 2005, Bielecki et al. \cite{BieleckiJinPliskaZhou} published the solution to this problem in a dynamic complete market setting.
 In both cases, the measure of risk of the portfolio is chosen as variance, and the risk-reward problem is understood as the ``Mean-Variance'' problem.  

 
Much research has been done in developing risk measures that focus on extreme events in the tail distribution where the portfolio loss occurs (variance does not differentiate loss or gain), and quantile-based models have thus far become the most popular choice. Among those, Conditional Value-at-Risk (CVaR) developed by Rockafellar and Uryasev \cite{RockafellarUryasevA} and  \cite{RockafellarUryasevB}, also known as Expected Shortfall by Acerbi and Tasche
\cite{AcerbiTasche},  has become a prominent candidate to replace variance in the portfolio selection problem. On the theoretical side, CVaR is a ``coherent risk measure'', a term coined by Artzner et al. \cite{ArtznerDelbaenEberHeath1} and \cite{ArtznerDelbaenEberHeath2} in pursuit of an axiomatic approach for defining properties that a `good' risk measure should possess.  On the practical side, the convex representation of CVaR from Rockafellar and Uryasev \cite{RockafellarUryasevA} opened the door of convex optimization for Mean-CVaR problem and gave it vast advantage in implementation.  In a one-period static setting, Rockafellar and Uryasev \cite{RockafellarUryasevA} demonstrated how linear programming can be used to solve the Mean-CVaR problem, making it a convincing alternative to the Markowitz \cite{Markowitz} Mean-Variance concept.

The work of Rockafellar and Uryasev \cite{RockafellarUryasevA} has raised huge interest for extending this approach.  Acerbi and Simonetti \cite{AcerbiSimonetti}, and Adam et al. \cite{AdamHoukariLaurent} generalized CVaR to spectral risk measure in a static setting.  Spectral risk measure is also known as Weighted Value-at-Risk (WVaR) by Cherny \cite{Cherny}, who in turn studied its optimization problem. Ruszczynsk and Shapiro \cite{RuszczynskiShapiro} revised CVaR into a multi-step dynamic risk measure, namely the ``conditional risk mapping for CVaR", and solved the corresponding Mean-CVaR problem using Rockafellar and Uryasev \cite{RockafellarUryasevA} technique for each time step.  When expected return is replaced by expected utility, the Utility-CVaR portfolio optimization problem is often studied in a continuous-time dynamic setting, see Gandy \cite{Gandy} and Zheng \cite{Zheng}.  More recently, the issue of robust implementation is dealt with in Quaranta and Zaffaroni \cite{QuarantaZaffaroni}, Gotoh et al. \cite{GotohShinozakiTakeda}, Huang et al. \cite{HuangZhuFabozziFukushima}, and El Karoui et al. \cite{ElKarouiLimVahn}.  Research on systemic risk that involves CVaR can be found in Acharya et al. \cite{AcharyaPedersenPhilipponRichardson}, Chen et al. \cite{ChenIyengarMoallemi}, and Adrian and Brunnermeier \cite{Adrian Brunnermeier}.

To the best of our knowledge, no complete characterization of solution has been done for the Mean-CVaR problem in a continuous-time dynamic setting.  Similar to Bielecki et al. \cite{BieleckiJinPliskaZhou}, we reduce the problem to a combination of a static optimization problem and a hedging problem with complete market assumption.  Our main contribution is that in solving the static optimization problem, we find a complete characterization whose nature is different than what is known in literature.  As a pure CVaR minimization problem without expected return constraint,  Sekine \cite{Sekine}, Li and Xu \cite{LiXuA}, Melnikov and Smirnov \cite{MelnikovSmirnov} found the optimal solution to be binary.  This is confirmed to be true for more general law-invariant risk (preference) measures minimization by Schied \cite{Schied}, and He and Zhou \cite{HeZhou}.  The key to finding the solution to be binary is the association of the Mean-CVaR problem to the Neyman-Pearson problem. We observe in Section \ref{Subsection: Outline} that the stochastic part of CVaR minimization can be transformed into Shortfall Risk minimization using the representation (CVaR is the Fenchel-Legendre dual of the Expected Shortfall) given by Rockafellar and Uryasev \cite{RockafellarUryasevA}.  F\"{o}llmer and Leukert \cite{FollmerLeukert} characterized the solution to the latter problem in a general semimartingale complete market model to be binary, where they have demonstrated its close relationship to the Neyman-Pearson problem of hypothesis testing between the risk neutral probability measure $\tilde{P}$ and the physical probability measure $P$. 

Adding the expected return constraint to WVaR minimization (CVaR is a particular case of WVaR), Cherny \cite{Cherny} found conditions under which the solution to the Mean-WVaR problem was still binary or nonexistent.  In this paper, we discuss all cases for solving the Mean-CVaR problem depending on a combination of two criteria: the level of the Radon-Nikod\'{y}m derivative $\frac{d\tilde{P}}{dP}$ relative to the confidence level of the risk measure; and the level of the return requirement. More specifically,  when the portfolio is uniformed bounded from above and below,  we find the optimal solution to be nonexistent or binary in some cases, and more interestingly, take three values in the most important case (see Case 4 of Theorem \ref{T: Solution to Step 2}).  When the portfolio is unbounded from above, in most cases (see Case 2 and 4 in Theorem \ref{T: unbounded above}), the solution is nonexistent, while portfolios of three levels still give sub-optimal solutions. Since the new solution we find can take not only the upper or the lower bound, but also a level in between, it can be viewed in part as a generalization of the binary solution for the Neyman-Pearson problem with an additional constraint on expectation.  



This paper is organized as follows. Section \ref{Section: Portfolio Selection} formulates the dynamic portfolio
selection problem, and compares the structure of the binary solution and the `three-level' solution, with an application of exact calculation in the Black-Scholes
model. Section \ref{Section: Solution} details the analytic solution in general where the proofs are delayed to Appendix \ref{Section: Appendix}; Section \ref{Section: Future Work} lists possible future work.

\section{The Structure of the Optimal Portfolio}\label{Section: Portfolio Selection}
\subsection{Main Problem}\label{Subsection: Outline}

Let  $(\Omega, \sF, (\sF)_{0\le t\le T}, P)$ be a filtered probability space that satisfies the usual conditions where $\sF_{0}$ is trivial and $\sF_{T}=\sF$. The market model consists of $d+1$ tradable assets: one riskless asset (money market account) and $d$ risky asset (stock).  Suppose the risk-free interest rate $r$ is a constant and the stock $S_{t}$ is a $d$-dimensional real-valued locally bounded semimartingale process.  Let the number of shares invested in the risky asset $\xi_{t}$ be a $d$-dimensional predictable process such that the stochastic integral with respect to $S_{t}$ is well-defined.  Then the value of a self-financing portfolio $X_{t}$  evolves according to the dynamics
\[dX_{t}=\xi_{t}dS_{t}+r(X_{t}-\xi_{t}S_{t})dt, \quad X_{0}=x_{0}.\]
Here $\xi_{t}dS_{t}$ and $\xi_{t}S_{t}$ are interpreted as inner products if the risky asset is multidimensional $d>1$. The portfolio selection problem is to find the best strategy $(\xi_{t})_{0\le t\le T}$ to minimize the Conditional Value-at-Risk (CVaR) of the final portfolio value $X_{T}$ at confidence level $0<\lambda<1$ , while requiring the expected value to remain above a constant $z$.\footnote{Krokhmal et al. \cite{KrokhmalPalmquistUryasev} showed conditions under which the problem of maximizing expected return with CVaR constraint is equivalent to the problem of minimizing CVaR with expected return constraint. In this paper, we use the term Mean-CVaR problem for both cases.}  In addition, we require uniform lower bound $x_{d}$ and upper bound $x_{u}$ on the value of the portfolio over time such that $-\infty<x_{d}<x_{0}<x_{u}\le\infty$.  Therefore, our \textbf{Main Dynamic Problem} is
\begin{align}\label{E: Main Problem Dynamic}
&\inf_{\xi_{t}}CVaR_{\lambda}(X_{T})\\
\text{subject to }\quad &E[X_{T}]\ge z,\quad x_{d}\le X_{t}\le x_{u}\,\,a.s. \quad\forall t\in[0,T].\notag
\end{align}
Note that the no-bankruptcy condition can be imposed by setting the lower bound to be $x_{d}=0$, and the portfolio value can be unbounded
from above by taking the upper bound as $x_{u}=\infty$.  Our solution will be based on the following complete market assumption.
\begin{assumption}\label{A: Complete Market and Continuous RND}
There is No Free Lunch with Vanishing Risk (as defined in Delbaen and Schachermayer \cite{DelbaenSchachermayer1}) and the market model is complete with a unique equivalent local martingale measure $\tilde{P}$ such that the Radon-Nikod\'{y}m derivative $\frac{d\tilde{P}}{dP}$ has a continuous distribution.
\end{assumption}
Under the above assumption any $\sF$-measurable random variable can be replicated by a dynamic portfolio. Thus the dynamic optimization problem (\ref{E: Main Problem Dynamic}) can be reduced to: first find the optimal solution $X^{**}$ to the \textbf{Main Static Problem},
\begin{align}\label{E: Main Problem Static}
&\inf_{X\in\sF}CVaR_{\lambda}(X)\\
\text{subject to }\quad &E[X]\ge z, \quad \tilde{E}[X]=x_{r}, \quad x_{d}\le X\le x_{u}\,\,a.s.\notag
\end{align}
if it exists, and then find the dynamic strategy that replicates the $\sF$-measurable random variable $X^{**}$.
Here the expectations $E$ and $\tilde{E}$ are taken under the physical probability measure $P$ and the risk neutral probability measure $\tilde{P}$ respectively. Constant $x_{r}=x_{0}e^{rT}$ is assumed to satisfy $-\infty<x_{d}<x_{0}\le x_{r}<x_{u}\le\infty$ and the additional capital constraint $\tilde{E}[X]=x_{r}$ is the key to make sure that the optimal solution can be replicated by a dynamic self-financing strategy with initial capital $x_{0}$.

Using the equivalence between Conditional Value-at-Risk and the Fenchel-Legendre dual of the Expected Shortfall derived in Rockafellar and Uryasev \cite{RockafellarUryasevA},
\begin{equation}\label{D: CVaR}
CVaR_{\lambda}(X)=\frac{1}{\lambda}\inf_{x\in\R}\left(E[(x-X)^{+}]-\lambda x\right), \quad \forall \lambda\in(0,1),
\end{equation}
the CVaR optimization problem (\ref{E: Main Problem Static}) can be reduced to an Expected Shortfall optimization problem which we name as the
\textbf{Two-Constraint Problem:}
\begin{description}
\item \textbf{Step 1:} Minimization of Expected Shortfall
\begin{align}\label{E: Step 1}
&v(x)=\inf_{X\in\sF}E[(x-X)^{+}]\\
\text{subject to }\quad &E[X]\ge z, \quad(\textit{return constraint})\notag\\
 &\tilde{E}[X]=x_{r}, \quad(\textit{capital constraint})\notag\\
 &x_{d}\le X\le x_{u}\,\,a.s.\notag
\end{align}
\item \textbf{Step 2:} Minimization of Conditional Value-at-Risk
\begin{equation}\label{E: Step 2}
\inf_{X\in\sF}CVaR_{\lambda}(X)=\frac{1}{\lambda}\inf_{x\in\R}\left(v(x)-\lambda x\right).
\end{equation}
\end{description}

To compare our solution to existing ones in literature, we also name an auxiliary problem which simply minimizes Conditional Value-at-Risk without the return constraint as the \textbf{One-Constraint Problem}: Step 1 in (\ref{E: Step 1}) is replaced by
\begin{description}
\item \textbf{Step 1:} Minimization of Expected Shortfall
\begin{align}\label{E: Step 1 One Constraint}
&v(x)=\inf_{X\in\sF}E[(x-X)^{+}]\\
\text{subject to }\quad
 &\tilde{E}[X]=x_{r}, \quad(\textit{capital constraint})\notag\\
 &x_{d}\le X\le x_{u}\,\,a.s.\notag
\end{align}
\end{description}
Step 2 in (\ref{E: Step 2}) remains the same.

\subsection{Main Result}\label{Subsection: Result}

This subsection is devoted to a conceptual comparison between the solutions to the {\em One-Constraint Problem} and the {\em Two-Constraint Problem}.
The solution to the Expected Shortfall Minimization problem in \textbf{Step 1} of the
\textbf{One-Constraint Problem}  is found by F\"{o}llmer and Leukert \cite{FollmerLeukert}  under Assumption \ref{A: Complete Market and Continuous RND}
to be binary in nature:
\begin{equation}\label{E: optimal X without expectation constraint}
X(x)=x_{d}\I_{A}+x\I_{A^{c}},
\quad\text{for } x_{d}<x<x_{u},
\end{equation}
 where $\I_{\cdot}(\omega)$ is the indicator function and set $A$ is defined as the collection of states where the Radon-Nikod\'{y}m derivative is above a threshold $\left\{\omega\in\Omega\,:\,\tfrac{d\tilde{P}}{dP}(\omega)>a\right\}$. 
This particular structure where  the optimal solution $X(x)$ takes only two values, namely the lower bound $x_{d}$ and $x$, is intuitively clear once the problems of  minimizing Expected Shortfall  and hypothesis testing between $P$ and $\tilde{P}$ are connected in F\"{o}llmer and Leukert \cite{FollmerLeukert}, the later being well-known to possess a binary solution by Neyman-Pearson Lemma.  There are various ways to prove the optimality.  Other than the Neyman-Pearson approach, it can be viewed as the solution from a convex duality perspective, see Theorem 1.19 in Xu \cite{Xu}. In addition, a simplified version to the proof of Proposition \ref{P: Three-Line Optimal for Step 2 in Two-Constraint Case} gives a  direct method using Lagrange multiplier for convex optimization. 

The solution to {\em Step 2} of {\em One-Constraint Problem}, and thus to the {\em Main Problems} in (\ref{E: Main Problem Dynamic}) and (\ref{E: Main
Problem Static}) as a pure risk minimization problem without the return constraint is given in Schied \cite{Schied}, Sekine \cite{Sekine}, and Li
and Xu \cite{LiXuA}.  Since Step 2 only involves minimization over a real-valued number $x$, the binary structure is preserved through this step.  
Under some technical conditions, the solution to
\textbf{Step 2} of the \textbf{One-Constraint Problem} is shown by Li and Xu \cite{LiXuA} (Theorem 2.10 and Remark 2.11) to be
\begin{align}
X^{*} &=x_{d}\I_{A^{*}}+x^{*}\I_{A^{*c}}, \quad\textbf{(Two-Line Configuration)}\label{E: optimal CVaR without expectation constraint}\\
CVaR_{\lambda}(X^{*}) &=-x_{r}+\frac{1}{\lambda}(x^{*}-x_{d})\left(P(A^{*})-\lambda \tilde{P}(A^{*})\right),\label{E: min CVaR two line}
\end{align}
where $(a^{*}, x^{*})$ is the solution to the
\textit{capital constraint} ($\tilde{E}[X(x)]=x_{r}$)  in {\em Step 1} and the \textit{first order Euler condition} ($v'(x)=0$) in
{\em Step 2}:
\begin{align}
x_{d}\tilde{P}(A)+x\tilde{P}(A^{c}) &=x_{r},  \quad(\textit{capital constraint})\label{E: capital constraint two line}\\
P(A)+\frac{\tilde{P}(A^{c})}{a}-\lambda &=0.  \quad(\textit{first order Euler condition})\label{E: euler condition two line}
\end{align}
A static portfolio holding only the riskless asset will yield a constant portfolio value $X\equiv x_{r}$ with $CVaR(X)=-x_{r}$.  The diversification by managing dynamically the exposure to risky assets decreases the risk of the overall portfolio by an amount shown in (\ref{E: min CVaR two line}). 
One interesting observation is that the optimal portfolio exists regardless whether the upper bound on the portfolio is finite $x_{u}<\infty$ or not $x_{u}=\infty$.  This conclusion will change drastically as we add the return constraint to the optimization problem.
The main result of this paper is to show that the optimal solution to the  {\em Two-Constraint Problem}, and thus the \textbf{Main Problem} (\ref{E: Main Problem Dynamic}) and (\ref{E: Main Problem Static}), does not have a  Neyman-Pearson type of binary solution, which we call {\em Two-Line Configuration} in (\ref{E: optimal CVaR without expectation constraint}); instead, it has a {\em Three-Line Configuration}.  Proposition \ref{P: Three-Line Optimal for Step 2 in Two-Constraint Case} and Theorem \ref{T: Solution to Step 2} prove that, when the upper bound is finite $x_{u}<\infty$ and under some technical conditions, the solution to \textbf{Step 2} of the \textbf{Two-Constraint Problem} turns out to be 
\begin{align}\label{E: optimal CVaR with expectation constraint}
X^{**} &=x_{d}\I_{A^{**}}+x^{**}\I_{B^{**}}+x_{u}\I_{D^{**}},  \quad\textbf{(Three-Line Configuration)}\\
CVaR_{\lambda}(X^{**}_{T}) &=\frac{1}{\lambda}\left((x^{**}-x_{d})P(A^{**})-\lambda x^{**}\right),
\end{align}
where  $(a^{**}, b^{**}, x^{**})$ is the solution to the \textit{capital constraint} and the \textit{first order Euler condition}, plus the additional \textit{return constraint} ($E[X(x)]=z$):
\begin{align}
x_{d}P(A)+xP(B)+x_{u}P(D) &=z,   \quad(\textit{return constraint})\label{E: return constraint three line}\\
x_{d}\tilde{P}(A)+x\tilde{P}(B)+x_{u}\tilde{P}(D) &=x_{r},   \quad(\textit{capital constraint})\label{E: capital constraint three line}\\
P(A)+\frac{\tilde{P}(B)-bP(B)}{a-b}-\lambda &=0.   \quad(\textit{first order Euler condition})\label{E: euler condition three line}
\end{align}
The sets in equation (\ref{E: return constraint three line})-(\ref{E: euler condition three line}) are defined by different levels of the Radon-Nikod\'{y}m derivative:
\[A=\left\{\omega\in\Omega\,:\,\tfrac{d\tilde{P}}{dP}(\omega)>a\right\},\quad 
B=\left\{\omega\in\Omega\,:\,b\le\tfrac{d\tilde{P}}{dP}(\omega)\le a\right\}, \quad 
D=\left\{\omega\in\Omega\,:\, \tfrac{d\tilde{P}}{dP}(\omega)<b\right\}.\]

When the upper bound is infinite $x_{u}=\infty$, Theorem \ref{T: unbounded above} shows that the solution for the optimal portfolio is no longer a {\em Three-Line Configuration}.  It can be pure money market account investment ({\em One-Line}), binary ({\em Two-Line}), or very likely nonexist. In the last case, the infimum of the CVaR can still be computed, and a sequence of  {\em Three-Line Configuration} portfolios can be found with their CVaR converging to the infimum.



\subsection{Example: Exact Calculation in the Black-Scholes Model}\label{Subsection: Example}

We show the closed-form calculation of the {\em Three-Line Configuration} (\ref{E: optimal CVaR with expectation constraint})-(\ref{E: euler condition three line}), as well as the corresponding optimal dynamic strategy in the benchmark Black-Scholes Model. Suppose an agent is trading between a money market account with interest rate $r$ and one stock\footnote{It is straight-forward to generalize the calculation to multi-dimensional Black-Scholes model.  Since we provide in this paper an analytical solution to the static CVaR minimization problem, calculation in other complete market models can be carried out as long as the dynamic hedge can be expressed in a simple manner.  
} that follows geometric Brownian motion $dS_{t}=\mu S_{t}dt+\sigma S_{t}dW_{t}$ with instantaneous rate of return $\mu$, volatility $\sigma$, and initial stock price $S_{0}$.  The endowment starts at $x_{0}$ and bankruptcy is prohibited at
any time,  $x_{d}=0$, before the final horizon $T$.  The expected terminal value $E[X_{T}]$ is required to be above a fixed level `$z$' to satisfy the return constraint.  When `$z$' is low, namely $z\le z^{*}\define E[X^{*}]$, where $X^{*}$  is the optimal portfolio (\ref{E: optimal CVaR without expectation constraint}) for the  {\em One Constraint Problem}, the return constraint is non-binding and obviously the {\em Two-Line Configuration} $X^{*}$ is optimal.  Let $\bar{z}$ be the highest expected value achievable by any self-financing portfolio starting with initial capital $x_{0}$ (see Definition \ref{D: zbar} and Lemma \ref{L: zbar is the highest achievable expected return}).  
When the return requirement becomes meaningful, i.e., $z\in(z^{*},\bar{z}]$, the {\em Three-Line Configuration} $X^{**}$ in (\ref{E: optimal CVaR with expectation constraint}) becomes optimal.

Since the Radon-Nikod\'{y}m derivative $\frac{d\tilde{P}}{dP}$ is a scaled power function of the final stock price which has a log-normal distribution, the probabilities in equations (\ref{E: capital constraint three line})-(\ref{E: euler condition three line}) can be computed in closed-form:
\begin{align*}
    P(A) &= N(-\tfrac{\theta\sqrt{T}}{2} -
    \tfrac{\ln a}{\theta\sqrt{T}}),\quad
    P(D) = 1-N(-\tfrac{\theta\sqrt{T}}{2} -
    \tfrac{\ln b}{\theta\sqrt{T}}), \quad
    P(B) = 1-P(A)-P(D),\notag\\
    \tilde{P}(A) &= N(\tfrac{\theta\sqrt{T}}{2} -
    \tfrac{\ln a}{\theta\sqrt{T}}),\quad
    \tilde{P}(D) = 1-N(\tfrac{\theta\sqrt{T}}{2} -
    \tfrac{\ln b}{\theta\sqrt{T}}),\quad
    \tilde{P}(B) = 1-\tilde{P}(A)-\tilde{P}(D),\notag
\end{align*}
where $\theta = \frac{\mu-r}{\sigma}$ and $N(\cdot)$ is the cumulative distribution function of a standard normal random variable.  From these, the solution $(a^{**}, b^{**}, x^{**})$ to equations (\ref{E: capital constraint three line})-(\ref{E: euler condition three line}) can be found numerically. The formulae for the dynamic value of the optimal portfolio $X_{t}^{**}$, the corresponding dynamic hedging strategy $ \xi_{t}^{**}$, and the associated final minimal risk $CVaR_{\lambda}(X^{**}_{T})$ are:
\begin{align*}
    X_{t}^{**} &= e^{-r(T-t)}[x^{**}N(d_{+}(a^{**},S_{t},t)) + x_{d}N(d_{-}(a^{**},S_{t},t))]\\
    &\qquad+e^{-r(T-t)}[x^{**}N(d_{-}(b^{**},S_{t},t)) + x_{u}N(d_{+}(b^{**},S_{t},t))]-e^{r(T-t)}x^{**},\\
    \xi_{t}^{**} &= \frac{x^{**}-x_{d}}{\sigma S_{t}\sqrt{2\pi(T-t)}}
    e^{-r(T-t)-\frac{d^{2}_{-}(a^{**},S_{t},t)}{2}}+\frac{x^{**}-x_{u}}{\sigma S_{t}\sqrt{2\pi(T-t)}}
    e^{-r(T-t)-\frac{d^{2}_{+}(b^{**},S_{t},t)}{2}},\\
    CVaR_{\lambda}(X^{**}_{T}) &=\frac{1}{\lambda}\left((x^{**}-x_{d})P(A^{**})-\lambda x^{**}\right),
\end{align*}    
where we define
$
    d_{-}(a,s,t) =\tfrac{1}{\theta\sqrt{T-t}}[-\ln a +
    \tfrac{\theta}{\sigma}(\tfrac{\mu+r-\sigma^{2}}{2}t-\ln\tfrac{s}{S_{0}}) +
    \tfrac{\theta^{2}}{2}(T-t)], \quad d_{+}(a,s,t) =-d_{-}(a,s,t).
$

Numerical results comparing the minimal risk for various levels of upper-bound $x_{u}$ and return constraint $z$ are summarized in Table \ref{table1}.  As expected the upper bound on the portfolio value $x_{u}$ has no impact on the {\em One-Constraint Problem}, as $(x^{*}, a^{*})$ and
$CVaR_{\lambda}(X^{*}_{T})$ are optimal whenever $x_{u}\ge x^{*}$. On contrary in  the {\em Two-Constraint Problem}, the stricter the return requirement $z$, the more the {\em Three-Line Configuration} $X^{**}$ deviates from the {\em Two-Line Configuration} $X^{*}$.  Stricter return requirement (higher $z$) implies higher minimal risk $CVaR_{\lambda}(X^{**}_{T})$; while less strict upper bound (higher $x_{u}$) decreases minimal risk $CVaR_{\lambda}(X^{**}_{T})$. Notably, under certain conditions in Theorem \ref{T: unbounded above}, for all levels of return $z\in(z^{*},\bar{z}]$, when $x_{u}\to\infty$, $CVaR_{\lambda}(X^{**}_{T})$ approaches $CVaR_{\lambda}(X^{*}_{T})$, as the optimal solution cease to exist in the limiting case.
\begin{table}[ht]
\begin{center}
{\small
\begin{tabular}{c|cc||c|ccc}
\multicolumn{3}{c||}{One-Constraint Problem} & \multicolumn{4}{|c}{Two-Constraint Problem}\\\hline
$x_{u}$ & 30 & 50 & $x_{u}$ & 30 & 30 & 50 \\
 & & & $z$ & 20 & 25 & 25\\\hline
$x^{*}$ & 19.0670 & 19.0670 & $x^{**}$ & 19.1258 & 19.5734 & 19.1434\\
$a^{*}$ & 14.5304 & 14.5304 & $a^{**}$ & 14.3765 & 12.5785 & 14.1677\\
 & & & $b^{**}$ & 0.0068 & 0.1326 & 0.0172\\\hline
$CVaR_{5\%}(X^{*}_{T})$ & -15.2118 & -15.2118 & $CVaR_{5\%}(X_{T}^{**})$ & -15.2067 & -14.8405 & -15.1483
\end{tabular}}
\caption{Black-Scholes example for One-Constraint (pure CVaR minimization) and Two-Constraint (Mean-CVaR optimization) problems with parameters:  $r=5\%$, $\mu =0.2,\,\sigma = 0.1$, $S_{0} = 10$, $T=2$, $x_{0} = 10$, $x_{d} = 0$, $\lambda=5\%$. Consequently, $z^{*}=18.8742$ and $\bar{z}=28.8866$.} 
\label{table1}
\end{center}
\end{table}

\begin{figure}[h]
 $$\includegraphics[scale=.6]{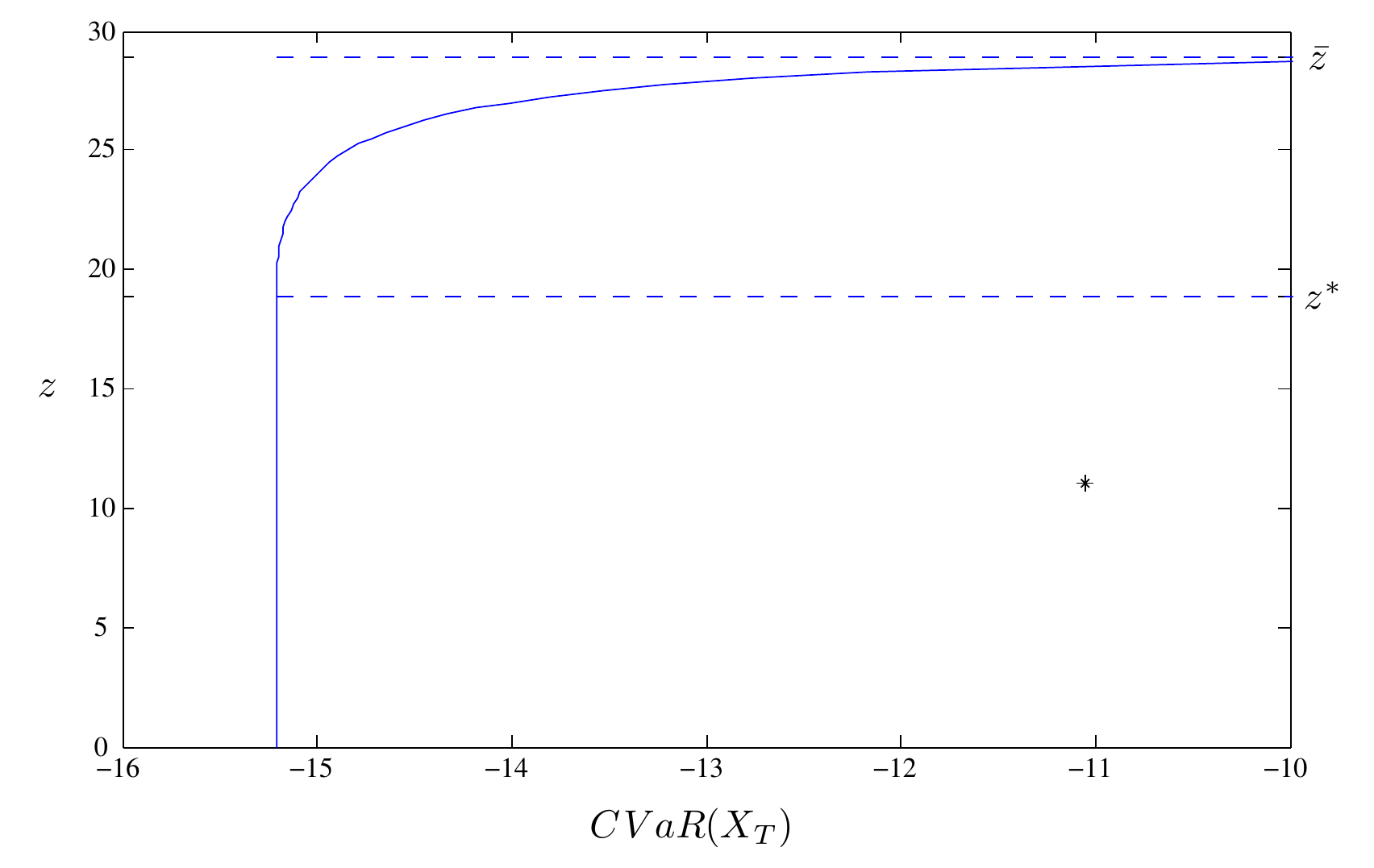}$$
 \caption{Efficient Frontier for Mean-CVaR Portfolio Selection} \label{graph1}
\end{figure}

Figure \ref{graph1} plots the efficient frontier of the above Mean-CVaR portfolio selection problem with fixed upper bound $x_{u} = 30$. The curve between return level $z^{*}$ and $\bar{z}$ are the Mean-CVaR efficient portfolio from various  {\em Three-Line Configurations}, while the straight line is the same Mean-CVaR efficient {\em Two-Line Configuration} when return constraint is non-binding.  
The star positioned at $(-x_{r},x_{r}) = (-11.0517,11.0517)$, where $x_{r} = x_{0}e^{rT}$, corresponds to the portfolio
that invests purely in the money market account. As a contrast to its position on the traditional 
Capital Market Line (the efficient frontier for a Mean-Variance portfolio selection problem), the
pure money market account portfolio is no longer efficient in the Mean-CVaR portfolio selection problem.

\section{Analytical Solution to the Portfolio Selection Problem}\label{Section: Solution}

Under Assumption \ref{A: Complete Market and Continuous RND},  the solution to the main Mean-CVaR optimization problem (\ref{E: Main Problem Static}), i.e., the {\em Two-Constraint Problem} (\ref{E: Step 1}) and (\ref{E: Step 2}), will be discussed in two separate cases where the upper bound for the portfolio value is finite or infinite. The main results are stated in Theorem \ref{T: Solution to Step 2} and Theorem \ref{T: unbounded above} respectively.
 To create a flow of showing clearly how the optimal solutions are related to the  {\em Two-Line} and {\em Three-Line Configurations}, all proofs  will be delayed to Appendix \ref{Section: Appendix}.  

\subsection{Case $x_{u}<\infty$: Finite Upper Bound }

We first define the general Three-Line Configuration and its degenerate Two-Line Configurations.
Recall from Section \ref{Subsection: Result} the definitions of the sets $A, B, D$ are
\begin{equation}\label{D: A, B, D definitions}
A=\left\{\omega\in\Omega\,:\,\tfrac{d\tilde{P}}{dP}(\omega)>a\right\},\quad
B=\left\{\omega\in\Omega\,:\,b\le\tfrac{d\tilde{P}}{dP}(\omega)\le a\right\},\quad
D=\left\{\omega\in\Omega\,:\, \tfrac{d\tilde{P}}{dP}(\omega)<b\right\}.
\end{equation}
\begin{definition}\label{D: Two definition A, B, D}Suppose $x\in[x_{d}, x_{u}]$.\hfill
\begin{enumerate}
\item Any \textbf{Three-Line Configuration} has the structure $X=x_{d}\I_{A}+x\I_{B}+x_{u}\I_{D}$.
\item The \textbf{Two-Line Configuration} $X=x\I_{B}+x_{u}\I_{D}$ is associated to the above definition in the case $a=\infty$,
$B=\left\{\omega\in\Omega\,:\,\tfrac{d\tilde{P}}{dP}(\omega)\ge b\right\}$ and
$D=\left\{\omega\in\Omega\,:\, \tfrac{d\tilde{P}}{dP}(\omega)<b\right\}$.\\
The \textbf{Two-Line Configuration} $X=x_{d}\I_{A}+x\I_{B}$ is associated to the above definition in the case\\ $b=0$,
$A=\left\{\omega\in\Omega\,:\,\tfrac{d\tilde{P}}{dP}(\omega)>a\right\}$, and
$B=\left\{\omega\in\Omega\,:\,\tfrac{d\tilde{P}}{dP}(\omega)\le a\right\}$.\\
The \textbf{Two-Line Configuration} $X=x_{d}\I_{A}+x_{u}\I_{D}$ is associated to the above definition in the case $a=b$,
$A=\left\{\omega\in\Omega\,:\,\tfrac{d\tilde{P}}{dP}(\omega)>a\right\}$, and
$D=\left\{\omega\in\Omega\,:\,\tfrac{d\tilde{P}}{dP}(\omega)< a\right\}$. 
\end{enumerate}
Moreover,
\begin{enumerate}
\item  \textbf{General Constraints} are the {\em capital constraint} and the equality part of the {\em expected return constraint}
for a \textbf{Three-Line Configuration} $X=x_{d}\I_{A}+x\I_{B}+x_{u}\I_{D}$:
\begin{align*}
&E[X]=x_{d}P(A)+xP(B)+x_{u}P(D)=z, \\
&\tilde{E}[X]=x_{d}\tilde{P}(A)+x\tilde{P}(B)+x_{u}\tilde{P}(D)=x_{r}.
\end{align*}
\item \textbf{Degenerated Constraints 1} are the {\em capital constraint} and the equality part of the {\em expected return constraint} for a \textbf{Two-Line Configuration} $X=x\I_{B}+x_{u}\I_{D}$:
\begin{align*}
&E[X]=xP(B)+x_{u}P(D)=z, \\
&\tilde{E}[X]=x\tilde{P}(B)+x_{u}\tilde{P}(D)=x_{r}.
\end{align*}
\textbf{Degenerated Constraints 2} are the {\em capital constraint} and the equality part of the {\em expected return constraint} for a \textbf{Two-Line Configuration} $X=x_{d}\I_{A}+x\I_{B}$:
\begin{align*}
&E[X]=x_{d}P(A)+xP(B)=z, \\
&\tilde{E}[X]=x_{d}\tilde{P}(A)+x\tilde{P}(B)=x_{r}.
\end{align*}
\textbf{Degenerated Constraints 3} are the {\em capital constraint} and the equality part of the {\em expected return constraint} for a \textbf{Two-Line Configuration} $X=x_{d}\I_{A}+x_{u}\I_{D}$:
\begin{align*}
&E[X]=x_{d}P(A)+x_{u}P(D)=z, \\
&\tilde{E}[X]=x_{d}\tilde{P}(A)+x_{u}\tilde{P}(D)=x_{r}.
\end{align*}
\end{enumerate}
Note that \textbf{Degenerated Constraints 1} correspond to the \textbf{General Constraints} when $a=\infty$;
\textbf{Degenerated Constraints 2} correspond to the \textbf{General Constraints} when $b=0$; and
\textbf{Degenerated Constraints 3} correspond to the \textbf{General Constraints} when $a=b$.
\end{definition}

We use the {\em Two-Line Configuration} $X=x_{d}\I_{A}+x_{u}\I_{D}$, where the value of the random variable $X$ takes either the upper or the lower bound, as well as its capital constraint to define the  `Bar-System' from which we calculate the highest achievable return.

\begin{definition}[The `Bar-System']\label{D: zbar}
For fixed $-\infty<x_{d}<x_{r}<x_{u}<\infty$, let $\bar{a}$ be a solution to the capital constraint $\tilde{E}[X]=x_{d}\tilde{P}(A)+x_{u}\tilde{P}(D)=x_{r}$  in \textbf{Degenerated Constraints 3} for the \textbf{Two-Line Configuration} $X=x_{d}\I_{A}+x_{u}\I_{D}$.  In the `Bar-System', $\bar{A}$, $\bar{D}$ and $\bar{X}$ are associated to the constant $\bar{a}$ in the sense  $\bar{X}=x_{d}\I_{\bar{A}}+x_{u}\I_{\bar{D}}$ where $\bar{A}=\left\{\omega\in\Omega\,:\,\tfrac{d\tilde{P}}{dP}(\omega)>\bar{a}\right\}$, and $\bar{D}=\left\{\omega\in\Omega\,:\,\tfrac{d\tilde{P}}{dP}(\omega)< \bar{a}\right\}$.   Define the expected return of the `Bar-System' as $\zbar=E[\bar{X}]=x_{d}P(\bar{A})+x_{u}P(\bar{D})$.
\end{definition}


\begin{lemma}\label{L: zbar is the highest achievable expected return}
$\zbar$ is the highest expected return that can be obtained  by  a self-financing portfolio with initial capital $x_{0}$ whose value is bounded between $x_{d}$ and $x_{u}$:
\[\zbar=\max_{X\in\sF}E[X]\quad s.t.\quad \tilde{E}[X]=x_{r}=x_{0}e^{rT}, \quad x_{d}\le X\le x_{u}\,\, a.s.\]
\end{lemma}

In the following lemma, we vary the `$x$' value in the {\em Two-Line Configurations} $X=x\I_{B}+x_{u}\I_{D}$ and $X=x_{d}\I_{A}+x\I_{B}$, while maintaining the capital constraints respectively. We observe their expected returns to vary between values $x_{r}$ and $\bar{z}$ in a monotone and continuous fashion.
 

\begin{lemma}\label{L: z decreases when x is between xd and xr and increases when x is between xr and xu}
For fixed $-\infty<x_{d}<x_{r}<x_{u}<\infty$. 
\begin{enumerate}
\item Given any $x\in[x_{d}, x_{r}]$, let `$b$' be a solution to the capital constraint $\tilde{E}[X]=x\tilde{P}(B)+x_{u}\tilde{P}(D)=x_{r}$  in \textbf{Degenerated Constraints 1} for the \textbf{Two-Line Configuration} $X=x\I_{B}+x_{u}\I_{D}$. Define the expected return of the resulting Two-Line Configuration as $z(x)=E[X]=xP(B)+x_{u}P(D)$.\footnote{Threshold `$b$' and consequently sets `$B$' and `$D$' are all dependent on `$x$' through the capital constraint, therefore $z(x)$ is not a linear function of $x$.}  Then $z(x)$ is a continuous function of $x$ and decreases from $\zbar$ to $x_{r}$ as $x$ increases from $x_{d}$ to $x_{r}$. 
\item Given any $x\in[x_{r}, x_{u}]$, let `$a$' be a solution to the capital constraint $\tilde{E}[X]=x_{d}\tilde{P}(A)+x\tilde{P}(B)=x_{r}$  in \textbf{Degenerated Constraints 2} for the \textbf{Two-Line Configuration} $X=x_{d}\I_{A}+x\I_{B}$. Define the expected return of the resulting Two-Line Configuration as $z(x)=E[X]=x_{d}P(A)+xP(B)$.  Then $z(x)$ is a continuous function of $x$ and increases from $x_{r}$ to $\zbar$ as $x$ increases from $x_{r}$ to $x_{u}$. 
\end{enumerate}
\end{lemma}

From now on, we will concern ourselves with requirements on the expected return in the interval $z\in[x_{r}, \zbar]$ because on one side Lemma \ref{L: zbar is the highest achievable expected return} ensures that there are no feasible solutions to the Main Problem (\ref{E: Main Problem Static}) if we require an expected return higher than $\zbar$. On the other side, 
Lemma \ref{L: zbar is the highest achievable expected return}, Lemma \ref{L: z decreases when x is between xd and xr and increases when x is between xr and xu} and Theorem \ref{T: Two Line Optimal in One Constraint Sase}  lead to the conclusion that a return constraint where $z\in(-\infty, x_{r})$ is too weak to differentiate the \textbf{Two-Constraint Problem} from the \textbf{One-Constraint Problem} as their optimal solutions concur.

\begin{definition}\label{D: xz1 and xz2}
For fixed $-\infty<x_{d}<x_{r}<x_{u}<\infty$, and a fixed level $z\in[x_{r}, \zbar]$, define $x_{z1}$ and $x_{z2}$ to be the corresponding $x$ value for \textbf{Two-Line Configurations} $X=x\I_{B}+x_{u}\I_{D}$ and $X=x_{d}\I_{A}+x\I_{B}$ that satisfy \textbf{Degenerated Constraints 1} and  \textbf{Degenerated Constraints 2} respectively.
\end{definition}

Definition \ref{D: xz1 and xz2} implies when we fix the level of expected return $z$, we can find two particular feasible solutions: $X=x_{z1}\I_{B}+x_{u}\I_{D}$ satisfying $\tilde{E}[X]=x_{z1}\tilde{P}(B)+x_{u}\tilde{P}(D)=x_{r}$ and $E[X]=x_{z1}P(B)+x_{u}P(D)=z$; $X=x_{d}\I_{A}+x_{z2}\I_{B}$ satisfying $\tilde{E}[X]=x_{d}\tilde{P}(A)+x_{z2}\tilde{P}(B)=x_{r}$ and $E[X]=x_{d}P(A)+x_{z2}P(B)=z$.  The values $x_{z1}$ and $x_{z2}$ are well-defined because Lemma \ref{L: z decreases when x is between xd and xr and increases when x is between xr and xu} guarantees $z(x)$ to be an invertible function in both cases.  We summarize in the following lemma whether the {\em Two-Line Configurations} satisfying the capital constraints meet or fail the return constraint as $x$ ranges over its domain $[x_{d}, x_{u}]$ for the {\em Two-Line} and {\em Three-Line Configurations} in Definition \ref{D: Two definition A, B, D}.


\begin{lemma}\label{L: Two-Line and their returns above or below z}
For fixed $-\infty<x_{d}<x_{r}<x_{u}<\infty$, and a fixed level $z\in[x_{r}, \zbar]$.
\begin{enumerate}
\item If we fix $x\in[x_{d}, x_{z1}]$, the Two-Line Configuration $X=x\I_{B}+x_{u}\I_{D}$ which satisfies the capital constraint $\tilde{E}[X]=x\tilde{P}(B)+x_{u}\tilde{P}(D)=x_{r}$ in Degenerated Constraints 1 \textbf{satisfies} the expected return constraint: $E[X]=xP(B)+x_{u}P(D)\ge z$;
\item If we fix $x\in(x_{z1}, x_{r}]$, the Two-Line Configuration $X=x\I_{B}+x_{u}\I_{D}$ which satisfies the capital constraint $\tilde{E}[X]=x\tilde{P}(B)+x_{u}\tilde{P}(D)=x_{r}$ in Degenerated Constraints 1 \textbf{fails} the expected return constraint: $E[X]=xP(B)+x_{u}P(D)<z$;
\item If we fix $x\in[x_{r}, x_{z2})$, the Two-Line Configuration $X=x_{d}\I_{A}+x\I_{B}$ which satisfies the capital constraint $\tilde{E}[X]=x_{d}\tilde{P}(A)+x\tilde{P}(B)=x_{r}$ in Degenerated Constraints 2 \textbf{fails} the expected return constraint: $E[X]=xP(B)+x_{u}P(D)<z$;
\item If we fix $x\in[x_{z2}, x_{u}]$, the Two-Line Configuration $X=x_{d}\I_{A}+x\I_{B}$ which satisfies the capital constraint $\tilde{E}[X]=x_{d}\tilde{P}(A)+x\tilde{P}(B)=x_{r}$ in Degenerated Constraints 2 \textbf{satisfies} the expected return constraint: $E[X]=xP(B)+x_{u}P(D)\ge z$.
\end{enumerate}
\end{lemma}

We turn our attention to solving \textbf{Step 1} of the \textbf{Two-Constraint Problem} (\ref{E: Step 1}):\\
\textbf{Step 1:} Minimization of Expected Shortfall
\begin{align*}
&v(x)=\inf_{X\in\sF}E[(x-X)^{+}]\\
\text{subject to }\quad &E[X]\ge z, \quad(\textit{return constraint})\notag\\
 &\tilde{E}[X]=x_{r}, \quad(\textit{capital constraint})\notag\\
 &x_{d}\le X\le x_{u}\,\,a.s.\notag
\end{align*}

Notice that a solution is called for any given real number $x$, independent of the return level $z$ or capital level $x_{r}$.  
From Lemma \ref{L: Two-Line and their returns above or below z} and the fact that the {\em Two-Line Configurations} are optimal solutions to \textbf{Step 1} of the \textbf{One-Constraint Problem} (see Theorem \ref{T: Two Line Optimal in One Constraint Sase}), we can immediately draw the following conclusion.

\begin{proposition}\label{P: Two Line Optimal for Step 1 when x is between xd and xz1; xz2 and xu}
For fixed $-\infty<x_{d}<x_{r}<x_{u}<\infty$, and a fixed level $z\in[x_{r}, \zbar]$.
\begin{enumerate}
\item  If we fix $x\in[x_{d}, x_{z1}]$, then there exists a \textbf{Two-Line Configuration} $X=x\I_{B}+x_{u}\I_{D}$ which is the optimal solution to \textbf{Step 1} of the \textbf{Two-Constraint Problem}; 
\item  If we fix $x\in[x_{z2}, x_{u}]$, then there exists a \textbf{Two-Line Configuration} $X=x_{d}\I_{A}+x\I_{B}$ which is the optimal solution to \textbf{Step 1} of the \textbf{Two-Constraint Problem}.
\end{enumerate}
\end{proposition}

When $x\in(x_{z1}, x_{z2})$, Lemma \ref{L: Two-Line and their returns above or below z} shows that the Two-Line Configurations which satisfy the capital constraints ($\tilde{E}[X]=x_{r}$) do not generate high enough expected return ($E[X]<z$) to be feasible anymore. It turns out that a novel solution of {\em Three-Line Configuration} is the answer: it can be shown to be both feasible and optimal.

\begin{lemma}\label{L: z decreases for Three-Line Configuration}
For fixed $-\infty<x_{d}<x_{r}<x_{u}<\infty$, and a fixed level $z\in[x_{r}, \zbar]$.
Given any $x\in(x_{z1}, x_{z2})$, let the pair of numbers $(a, b)\in\R^{2}$  ($b\le a$) be a solution to the capital constraint $\tilde{E}[X]=x_{d}\tilde{P}(A)+x\tilde{P}(B)+x_{u}\tilde{P}(D)=x_{r}$ in \textbf{General Constraints} for the \textbf{Three-Line Configuration} $X=x_{d}\I_{A}+x\I_{B}+x_{u}\I_{D}$.  Define the expected return of the resulting  Three-Line Configuration as $z(a,b)=E[X]=x_{d}P(A)+xP(B)+x_{u}P(D)$.  Then $z(a, b)$ is a continuous function which decreases from $\zbar$ to a number below $z$: 
\begin{enumerate}
\item When $a=b=\bar{a}$ from Definition \ref{D: zbar} of `Bar-System',  the Three-Line Configuration degenerates to $X=\bar{X}$ and $z(\bar{a},\bar{a})=E[\bar{X}]=\zbar$.  
\item When $b<\bar{a}$ and $a>\bar{a}$, $z(a, b)$ decreases continuously as $b$ decreases and $a$ increases.
\item In the extreme case when $a=\infty$, the Three-Line configuration becomes the Two-Line Configuration $X=x\I_{B}+x_{u}\I_{D}$; in the extreme case when $b=0$, the Three-Line configuration becomes the Two-Line Configuration $X=x_{d}\I_{A}+x\I_{B}$.  In either case, the expected value is below $z$ by Lemma \ref{L: Two-Line and their returns above or below z}.
\end{enumerate}
\end{lemma}

\begin{proposition}\label{P: Three Line Optimal for Step 1 when x is between xz1 and xz2}
For fixed $-\infty<x_{d}<x_{r}<x_{u}<\infty$, and a fixed level $z\in[x_{r}, \zbar]$.  If we fix $x\in(x_{z1}, x_{z2})$, then there exists a \textbf{Three-Line Configuration} $X=x_{d}\I_{A}+x\I_{B}+x_{u}\I_{D}$ that satisfies the \textbf{General Constraints} which is the optimal solution to \textbf{Step 1} of the \textbf{Two-Constraint Problem}.
\end{proposition}

Combining Proposition \ref{P: Two Line Optimal for Step 1 when x is between xd and xz1; xz2 and xu} and Proposition \ref{P: Three Line Optimal for Step 1 when x is between xz1 and xz2}, we arrive to the following result on the optimality of the Three-Line Configuration.

\begin{theorem}[Solution to \textbf{Step 1:} Minimization of Expected Shortfall]\label{T: Solution to Step 1}\hfil\break
For fixed $-\infty<x_{d}<x_{r}<x_{u}<\infty$, and a fixed level $z\in[x_{r}, \zbar]$.  $X(x)$ and the corresponding value function $v(x)$ described below are optimal solutions to \textbf{Step 1: Minimization of Expected Shortfall} of the \textbf{Two-Constraint Problem}:
\begin{itemize}
\item $x\in(-\infty, x_{d}]$:
$X(x) =$ any random variable with values in $[x_{d}, x_{u}]$ satisfying  both the capital constraint $\tilde{E}[X(x)]=x_{r}$ and the return constraint $E[X(x)]\ge z$.  $v(x) = 0.$
\item $x\in[x_{d}, x_{z1}]$:
$X(x) =$ any random variable with values in $[x, x_{u}]$ satisfying  both the capital constraint $\tilde{E}[X(x)]=x_{r}$ and the return constraint $E[X(x)]\ge z$. $v(x) = 0.$
\item $x\in(x_{z1}, x_{z2})$:
$X(x) =x_{d}\I_{A_{x}}+x\I_{B_{x}}+x_{u}\I_{D_{x}}$ where $A_{x}, B_{x}, D_{x}$ are determined by $a_{x}$ and $b_{x}$ as in (\ref{D: A, B, D definitions}) satisfying  the General Constraints:
$\tilde{E}[X(x)]=x_{r}$ and $E[X(x)]=z$. $v(x) = (x-x_{d})P(A_{x}).$
\item $x\in[x_{z2}, x_{u}]$:
$X(x) =x_{d}\I_{A_{x}}+x\I_{B_{x}}$ where $A_{x}, B_{x}$ are determined by $a_{x}$ as in Definition \ref{D: Two definition A, B, D} satisfying  both the capital constraint $\tilde{E}[X(x)]=x_{r}$ and the return constraint $E[X(x)]\ge z$.
$v(x) = (x-x_{d})P(A_{x}).$
\item $x\in[x_{u}, \infty)$:
$X(x) =x_{d}\I_{\bar{A}}+x_{u}\I_{\bar{B}}=\bar{X}$ where $\bar{A}, \bar{B}$ are associated to $\bar{a}$ as in Definition \ref{D: zbar} satisfying  both the capital constraint $\tilde{E}[X(x)]=x_{r}$ and the return constraint $E[X(x)]=\zbar\ge z$. \\$v(x) = (x-x_{d})P(\bar{A})+(x-x_{u})P(\bar{B}).$
\end{itemize}
\end{theorem}

To solve \textbf{Step 2}  of the \textbf{Two-Constraint Problem} (\ref{E: Step 2}), and thus the Main Problem (\ref{E: Main Problem Static}), we need to minimize
\[\frac{1}{\lambda}\inf_{x\in\R}(v(x)-\lambda x),\] where  $v(x)$ has been computed in Theorem \ref{T: Solution to Step 1}.
Depending on the $z$ level in the return constraint being lenient or strict,  the solution is sometimes obtained by the Two-Line Configuration which is optimal to the One-Constraint Problem, and other times obtained by a true Three-Line configuration.  To proceed in this direction, we recall the solution to the \textbf{One-Constraint Problem} from Li and Xu \cite{LiXuA}.

\begin{theorem}[Theorem 2.10 and Remark 2.11 in Li and Xu \cite{LiXuA} when $x_{u}<\infty$]\label{T: Two Line Optimal in One Constraint Sase}\hfill
\begin{enumerate}
\item
Suppose $\esssup\frac{d\tilde{P}}{dP}\le\frac{1}{\lambda}$.  $X=x_{r}$  is the optimal solution to \textbf{Step 2: Minimization of Conditional Value-at-Risk} of the \textbf{One-Constraint Problem} and the associated minimal risk is
\[CVaR(X)=-x_{r}.\]
\item
Suppose $\esssup\frac{d\tilde{P}}{dP}>\frac{1}{\lambda}$.
\begin{itemize}
\item If $\frac{1}{\bar{a}}\le\frac{\lambda-P(\bar{A})}{1-\tilde{P}(\bar{A})}$ (see Definition \ref{D: zbar} for the `Bar-System'), then $\bar{X}=x_{d}\I_{\bar{A}}+x_{u}\I_{\bar{D}}$ is the optimal solution to \textbf{Step 2: Minimization of Conditional Value-at-Risk} of the \textbf{One-Constraint Problem} and the associated minimal risk is
\[CVaR(\bar{X})=-x_{r}+\frac{1}{\lambda}(x_{u}-x_{d})(P(\bar{A})-\lambda\tilde{P}(\bar{A})).\]
\item Otherwise, let $a^{*}$ be the solution to the equation $\frac{1}{a}=\frac{\lambda-P(A)}{1-\tilde{P}(A)}$.  Associate sets $A^{*}=\left\{\omega\in\Omega\,:\,\frac{d\tilde{P}}{dP}(\omega)>a^{*}\right\}$ and $B^{*}=\left\{\omega\in\Omega\,:\,\frac{d\tilde{P}}{dP}(\omega)\le a^{*}\right\}$ to level $a^{*}$.  Define $x^{*}=\frac{x_{r}-x_{d}\tilde{P}(A^{*})}{1-\tilde{P}(A^{*})}$ so that configuration \[X^{*}=x_{d}\I_{A^{*}}+x^{*}\I_{B^{*}}\] satisfies the capital constraint $\tilde{E}[X^{*}]=x_{d}\tilde{P}(A^{*})+x^{*}\tilde{P}(B^{*})=x_{r}$.\footnote{Equivalently, $(a^{*}, x^{*})$ can be viewed as the solution to the capital constraint and the first order Euler condition in equations (\ref{E: capital constraint two line}) and (\ref{E: euler condition two line}).}  Then $X^{*}$ (we call the {\bf`Star-System'}) is the optimal solution to \textbf{Step 2: Minimization of Conditional Value-at-Risk} of the \textbf{One-Constraint Problem} and the associated minimal risk is
\[CVaR(X^{*})=-x_{r}+\frac{1}{\lambda}(x^{*}-x_{d})(P(A^{*})-\lambda\tilde{P}(A^{*})).\]
\end{itemize}
\end{enumerate}
\end{theorem}

\begin{definition}\label{D: z*}
In part 2 of Theorem \ref{T: Two Line Optimal in One Constraint Sase}, define $z^{*}=\zbar$ in the first case when $\frac{1}{\bar{a}}\le\frac{\lambda-P(\bar{A})}{1-\tilde{P}(\bar{A})}$; define $z^{*}=E[X^{*}]$ in the second case when $\frac{1}{\bar{a}}>\frac{\lambda-P(\bar{A})}{1-\tilde{P}(\bar{A})}$.
\end{definition}

We see that when $z$ is smaller than $z^{*}$, the binary solutions $X^{*}$ and $\bar{X}$ provided in Theorem \ref{T: Two Line Optimal in One Constraint Sase} are indeed the optimal solutions to Step 2 of the Two-Constraint Problem.  However, when $z$ is greater than $z^{*}$, these Two-Line Configurations are no longer feasible in the Two-Constraint Problem. We now show that the Three-Line Configuration is not only feasible but also optimal.  First we establish the convexity of the objective function and its continuity in a Lemma.

\begin{lemma}\label{L: convexity of v(x)}
$v(x)$ is a convex function for $x\in\R$, and thus continuous.
\end{lemma}

\begin{proposition}\label{P: Three-Line Optimal for Step 2 in Two-Constraint Case}
For fixed $-\infty<x_{d}<x_{r}<x_{u}<\infty$, and a fixed level $z\in(z^{*}, \zbar]$.\\ Suppose $\esssup\frac{d\tilde{P}}{dP}>\frac{1}{\lambda}$.    The solution $(a^{**}, b^{**},x^{**})$ (and consequently, $A^{**}, B^{**}$ and $D^{**}$)  to the equations
\begin{align*}
x_{d}P(A)+xP(B)+x_{u}P(D) &=z, \quad (\text{return constraint}) \\
x_{d}\tilde{P}(A)+x\tilde{P}(B)+x_{u}\tilde{P}(D) &=x_{r}, \quad (\text{capital constraint})\\
P(A)+\frac{\tilde{P}(B)-bP(B)}{a-b}-\lambda &=0, \quad (\text{first order Euler condition})
\end{align*}
exists.  $X^{**}=x_{d}\I_{A^{**}}+x^{**}\I_{B^{**}}+x_{u}\I_{D^{**}} $ (we call the {\bf`Double-Star System'})  is the optimal solution to \textbf{Step 2: Minimization of Conditional Value-at-Risk} of the \textbf{Two-Constraint Problem} 
and the associated minimal risk is
\[CVaR(X^{**})=\frac{1}{\lambda}\left((x^{**}-x_{d})P(A^{**})-\lambda x^{**}\right).\]
\end{proposition}

Putting together Proposition \ref{P: Three-Line Optimal for Step 2 in Two-Constraint Case} with Theorem \ref{T: Two Line Optimal in One Constraint Sase}, we arrive to the {\bf Main Theorem} of this paper.

\begin{theorem}[Minimization of Conditional Value-at-Risk When $x_{u}<\infty$]\label{T: Solution to Step 2}\hfil\break
For fixed $-\infty<x_{d}<x_{r}<x_{u}<\infty$.
\begin{enumerate}
\item Suppose $\esssup\frac{d\tilde{P}}{dP}\le\frac{1}{\lambda}$ and $z=x_{r}$.  The pure money market account investment $X=x_{r}$  is the optimal solution to \textbf{Step 2: Minimization of Conditional Value-at-Risk} of the \textbf{Two-Constraint Problem} and the associated minimal risk is
\[CVaR(X)=-x_{r}.\]

\item Suppose $\esssup\frac{d\tilde{P}}{dP}\le\frac{1}{\lambda}$ and $z\in(x_{r}, \zbar]$.  The optimal solution to \textbf{Step 2: Minimization of Conditional Value-at-Risk} of the \textbf{Two-Constraint Problem} does not exist and the minimal risk is
\[CVaR(X)=-x_{r}.\]

\item Suppose $\esssup\frac{d\tilde{P}}{dP}>\frac{1}{\lambda}$  and $z\in[x_{r}, z^{*}]$ (see Definition \ref{D: z*} for $z^{*}$).
\begin{itemize}
\item If $\frac{1}{\bar{a}}\le\frac{\lambda-P(\bar{A})}{1-\tilde{P}(\bar{A})}$ (see Definition \ref{D: zbar}), then the {\bf `Bar-System'} $\bar{X}=x_{d}\I_{\bar{A}}+x_{u}\I_{\bar{D}}$ is the optimal solution to \textbf{Step 2: Minimization of Conditional Value-at-Risk} of the \textbf{Two-Constraint Problem} and the associated minimal risk is
\[CVaR(\bar{X})=-x_{r}+\frac{1}{\lambda}(x_{u}-x_{d})(P(\bar{A})-\lambda\tilde{P}(\bar{A})).\]
\item Otherwise, the {\bf `Star-System'} $X^{*}=x_{d}\I_{A^{*}}+x^{*}\I_{B^{*}}$ defined in Theorem \ref{T: Two Line Optimal in One Constraint Sase} is the optimal solution to \textbf{Step 2: Minimization of Conditional Value-at-Risk} of the \textbf{Two-Constraint Problem} and the associated minimal risk is
\[CVaR(X^{*})=-x_{r}+\frac{1}{\lambda}(x^{*}-x_{d})(P(A^{*})-\lambda\tilde{P}(A^{*})).\]
\end{itemize}

\item Suppose $\esssup\frac{d\tilde{P}}{dP}>\frac{1}{\lambda}$  and $z\in (z^{*}, \zbar]$.
the {\bf `Double-Star-Sytem'} $X^{**}=x_{d}\I_{A^{**}}+x^{**}\I_{B^{**}}+x_{u}\I_{D^{**}}$
defined in Proposition \ref{P: Three-Line Optimal for Step 2 in Two-Constraint Case} is the optimal solution to \textbf{Step 2: Minimization of Conditional Value-at-Risk} of the \textbf{Two-Constraint Problem} and the associated minimal risk is
\[CVaR(X^{**})=\frac{1}{\lambda}\left((x^{**}-x_{d})P(A^{**})-\lambda x^{**}\right).\]
\end{enumerate}

\end{theorem}

We observe that the pure money market account investment is rarely optimal.  When the Radon-Nikod\'{y}m derivative is bounded above by the reciprocal of the confidence level of the risk measure ($\esssup\frac{d\tilde{P}}{dP}\le\frac{1}{\lambda}$), a condition not satisfied in the Black-Scholes model, the solution does not exist unless the return requirement coincide with the risk-free rate.  When the Radon-Nikod\'{y}m derivative exceeds $\frac{1}{\lambda}$ with positive probability, and the return constraint is low $z\in[x_{r}, z^{*}]$, the Two-Line Configuration which is optimal to the $CVaR$ minimization problem without the return constraint is also the optimal to the Mean-CVaR problem.  However, in the more interesting case where the return constraint is materially high $z\in (z^{*}, \zbar]$, the optimal Three-Line-Configuration sometimes takes the value of the upper bound $x_{u}$ to raise the expected return at the cost the minimal risk will be at a higher level.  This analysis complies with the numerical example shown in Section \ref{Subsection: Example}.

\subsection{Case $x_{u}=\infty$: No Upper Bound}

We first restate the solution to the \textbf{One-Constraint Problem} from  Li and Xu \cite{LiXuA} in the current context: when $x_{u}=\infty$, where we interpret $\bar{A}=\Omega$ and $\zbar=\infty$.

\begin{theorem}[Theorem 2.10 and Remark 2.11 in Li and Xu \cite{LiXuA} when $x_{u}=\infty$]\label{T: Two Line Optimal in One Constraint Case Infinite Upper Bound}\hfill
\begin{enumerate}
\item
Suppose $\esssup\frac{d\tilde{P}}{dP}\le\frac{1}{\lambda}$.  The pure money market account investment $X=x_{r}$  is the optimal solution to \textbf{Step 2: Minimization of Conditional Value-at-Risk} of the \textbf{One-Constraint Problem} and the associated minimal risk is
\[CVaR(X)=-x_{r}.\]
\item
Suppose $\esssup\frac{d\tilde{P}}{dP}>\frac{1}{\lambda}$.
The {\bf `Star-System'} $X^{*}=x_{d}\I_{A^{*}}+x^{*}\I_{B^{*}}$ defined in Theorem \ref{T: Two Line Optimal in One Constraint Sase}  is the optimal solution to \textbf{Step 2: Minimization of Conditional Value-at-Risk} of the \textbf{One-Constraint Problem} and the associated minimal risk is
\[CVaR(X^{*})=-x_{r}+\frac{1}{\lambda}(x^{*}-x_{d})(P(A^{*})-\lambda\tilde{P}(A^{*})).\]
\end{enumerate}
\end{theorem}

We observe that although there is no upper bound for the portfolio value, the optimal solution remains bounded from above, and the minimal $CVaR$ is bounded from below.  The problem of purely minimizing $CVaR$ risk of a self-financing portfolio (bounded below by $x_{d}$ to exclude arbitrage) from initial capital $x_{0}$ is feasible in the sense that the risk will not approach $-\infty$ and the minimal risk is achieved by an optimal portfolio.   When we add substantial return constraint to the $CVaR$ minimization problem, although the minimal risk can still  be calculated in the most important case ({\em Case 4} in Theorem \ref{T: unbounded above}), it is truly an infimum and not a minimum, thus it can be approximated closely by a sub-optimal portfolio, but not achieved by an optimal portfolio. 

\begin{theorem}[Minimization of Conditional Value-at-Risk When $x_{u}=\infty$]\label{T: unbounded above}\hfil\break
For fixed $-\infty<x_{d}<x_{r}<x_{u}=\infty$.
\begin{enumerate}
\item Suppose $\esssup\frac{d\tilde{P}}{dP}\le\frac{1}{\lambda}$ and $z=x_{r}$.  The pure money market account investment $X=x_{r}$  is the optimal solution to \textbf{Step 2: Minimization of Conditional Value-at-Risk} of the \textbf{Two-Constraint Problem} and the associated minimal risk is
\[CVaR(X)=-x_{r}.\]

\item Suppose $\esssup\frac{d\tilde{P}}{dP}\le\frac{1}{\lambda}$ and $z\in(x_{r}, \infty)$.  The optimal solution to \textbf{Step 2: Minimization of Conditional Value-at-Risk} of the \textbf{Two-Constraint Problem} does not exist and the minimal risk is
\[CVaR(X)=-x_{r}.\]

\item Suppose $\esssup\frac{d\tilde{P}}{dP}>\frac{1}{\lambda}$  and $z\in[x_{r}, z^{*}]$.  The {\bf `Star-System'} $X^{*}=x_{d}\I_{A^{*}}+x^{*}\I_{B^{*}}$ defined in Theorem \ref{T: Two Line Optimal in One Constraint Sase}  is the optimal solution to \textbf{Step 2: Minimization of Conditional Value-at-Risk} of the \textbf{Two-Constraint Problem} and the associated minimal risk is
\[CVaR(X^{*})=-x_{r}+\frac{1}{\lambda}(x^{*}-x_{d})(P(A^{*})-\lambda\tilde{P}(A^{*})).\]

\item Suppose $\esssup\frac{d\tilde{P}}{dP}>\frac{1}{\lambda}$  and $z\in(z^{*}, \infty)$.
The optimal solution to \textbf{Step 2: Minimization of Conditional Value-at-Risk} of the \textbf{Two-Constraint Problem} does not exist and the minimal risk is
\[CVaR(X^{*})=-x_{r}+\frac{1}{\lambda}(x^{*}-x_{d})(P(A^{*})-\lambda\tilde{P}(A^{*})).\]

\end{enumerate}
\end{theorem}

\begin{remark}
From the proof of the above theorem in {\bf Appendix} \ref{Section: Appendix}, we note that in case {\it 4}, we can always find a Three-Line Configuration as a sub-optimal solution, i.e., there exists for every $\epsilon>0$, a corresponding portfolio $X_{\epsilon}=x_{d}\I_{A_{\epsilon}}+x_{\epsilon}\I_{B_{\epsilon}}+\alpha_{\epsilon}\I_{D_{\epsilon}}$ which satisfies the {\em General Constraints} and produces a $CVaR$ level close to the lower bound:
$CVaR(X_{\epsilon})\le CVaR(X^{*})+\epsilon.$
\end{remark}

\section{Future Work}\label{Section: Future Work}

The second part of Assumption \ref{A: Complete Market and Continuous RND}, namely the Radon-Nikod\'{y}m derivative $\frac{d\tilde{P}}{dP}$ having a continuous distribution, is imposed for the simplification it brings to the presentation in the main theorems.
Further work can be done when this assumption is weakened.  We expect that the main results should still hold, albeit in a more complicated form.\footnote{The outcome in its format resembles techniques employed in F\"{o}llmer and Leukert \cite{FollmerLeukert} and  Li and Xu \cite{LiXuA} where the point masses on the thresholds for the Radon-Nikod\'{y}m derivative in (\ref{D: A, B, D definitions}) have to be dealt with carefully.}  It will also be interesting to extend the closed-form solution for Mean-CVaR minimization by replacing CVaR with Law-Invariant Convex Risk Measures in general.  Another direction will be to employ dynamic risk measures into the current setting.

Although in this paper we focus on the complete market solution, to solve the problem in an incomplete market setting, the exact hedging argument via Martingale Representation Theorem that translates the dynamic problem (\ref{E: Main Problem Dynamic}) into the static problem (\ref{E: Main Problem Static}) has to be replaced by a super-hedging argument via Optional Decomposition developed by Kramkov \cite{Kramkov}, and F\"{o}llmer and Kabanov \cite{FollmerKabanov}. The detail is similar to the process carried out for Shortfall Risk Minimization in F\"{o}llmer and Leukert \cite{FollmerLeukert},  Convex Risk Minimization in Rudloff \cite{Rudloff}, and law-invariant risk preference in He and Zhou \cite{HeZhou}.  The curious question is: Will the Third-Line Configuration remain optimal?

\section{Appendix}\label{Section: Appendix}

\hfil\break {\sc Proof of Lemma \ref{L: zbar is the highest achievable expected return}.}  The problem of
\[\zbar=\max_{X\in\sF}E[X]\quad s.t.\quad \tilde{E}[X]=x_{r}, \quad x_{d}\le X\le x_{u}\,a.s.\] is equivalent to the Expected Shortfall Problem
\[\zbar=-\min_{X\in\sF}E[(x_{u}-X)^{+}]\quad s.t.\quad \tilde{E}[X]=x_{r}, \quad X\ge x_{d}\,a.s.\]
Therefore, the answer is immediate.
$\endproof$

\hfil\break {\sc Proof of Lemma \ref{L: z decreases when x is between xd and xr and increases when x is between xr and xu}.}  Choose $x_{d}\le x_{1}<x_{2}\le x_{r}$.
Let $X_{1}=x_{1}\I_{B_{1}}+x_{u}\I_{D_{1}}$ where $B_{1}=\left\{\omega\in\Omega\,:\,\tfrac{d\tilde{P}}{dP}(\omega)\ge b_{1}\right\}$ and $D_{1}=\left\{\omega\in\Omega\,:\, \tfrac{d\tilde{P}}{dP}(\omega)<b_{1}\right\}$.  Choose $b_{1}$ such that $\tilde{E}[X_{1}]=x_{r}$.  This capital constraint means $x_{1}\tilde{P}(B_{1})+x_{u}\tilde{P}(D_{1})=x_{r}$.  Since $\tilde{P}(B_{1})+\tilde{P}(D_{1})=1$, $\tilde{P}(B_{1})=\frac{x_{u}-x_{r}}{x_{u}-x_{1}}$ and $\tilde{P}(D_{1})=\frac{x_{r}-x_{1}}{x_{u}-x_{1}}$.  Define $z_{1}=E[X_{1}]$.  Similarly, $z_{2}, X_{2}, B_{2}, D_{2}, b_{2}$ corresponds to $x_{2}$ where $b_{1}>b_{2}$ and $\tilde{P}(B_{2})=\frac{x_{u}-x_{r}}{x_{u}-x_{2}}$  and $\tilde{P}(D_{2})=\frac{x_{r}-x_{2}}{x_{u}-x_{2}}$.  Note that $D_{2}\subset D_{1}$,  $B_{1}\subset B_{2}$ and $D_{1}\backslash D_{2}=B_{2}\backslash B_{1}$.  We have
\begin{align*}
z_{1}-z_{2} &= x_{1}P(B_{1})+x_{u}P(D_{1})-x_{2}P(B_{2})-x_{u}P(D_{2})\\
&=(x_{u}-x_{2})P(B_{2}\backslash B_{1})-(x_{2}-x_{1})P(B_{1})\\
&=(x_{u}-x_{2})P\left(b_{2}<\tfrac{d\tilde{P}}{dP}(\omega)< b_{1}\right)-(x_{2}-x_{1})P\left(\tfrac{d\tilde{P}}{dP}(\omega)\ge b_{1}\right)\\
&=(x_{u}-x_{2})\int_{\left\{b_{2}<\tfrac{d\tilde{P}}{dP}(\omega)< b_{1}\right\}}\tfrac{dP}{d\tilde{P}}(\omega)d\tilde{P}(\omega)
-(x_{2}-x_{1})\int_{\left\{\tfrac{d\tilde{P}}{dP}(\omega)\ge b_{1}\right\}}\tfrac{dP}{d\tilde{P}}(\omega)d\tilde{P}(\omega)\\
&>(x_{u}-x_{2})\frac{1}{b_{1}}\tilde{P}(B_{2}\backslash B_{1})-(x_{2}-x_{1})\frac{1}{b_{1}}\tilde{P}(B_{1})\\
&=(x_{u}-x_{2})\frac{1}{b_{1}}\left(\frac{x_{u}-x_{r}}{x_{u}-x_{2}}-\frac{x_{u}-x_{r}}{x_{u}-x_{1}}\right)
-(x_{2}-x_{1})\frac{1}{b_{1}}\frac{x_{u}-x_{r}}{x_{u}-x_{1}}=0.
\end{align*}
For any given $\epsilon>0$, choose $x_{2}-x_{1}\le \epsilon$, then
\begin{align*}
z_{1}-z_{2} &=(x_{u}-x_{1})P(B_{2}\backslash B_{1})-(x_{2}-x_{1})P(B_{2})\\
&\le (x_{u}-x_{1})P(B_{2}\backslash B_{1})\\
&\le (x_{u}-x_{1})\left(\frac{x_{u}-x_{r}}{x_{u}-x_{2}}-\frac{x_{u}-x_{r}}{x_{u}-x_{1}}\right)\\
&\le \frac{(x_{2}-x_{1})(x_{u}-x_{r})}{x_{u}-x_{2}}\le x_{2}-x_{1}\le \epsilon.
\end{align*}
Therefore, $z$ decreases continuously as $x$ increases when $x\in[x_{d}, x_{r}]$.  When $x=x_{d}$, $z=\zbar$ from Definition \ref{D: zbar}.  When $x=x_{r}$,  $X\equiv x_{r}$ and $z=E[X]=x_{r}$.  Similarly, we can show that $z$ increases continuously from $x_{r}$ to $\zbar$ as $x$ increases from $x_{r}$ to $x_{u}$.
$\endproof$

\hfil\break  Lemma \ref{L: Two-Line and their returns above or below z} is a logical consequence of Lemma \ref{L: z decreases when x is between xd and xr and increases when x is between xr and xu} and Definition \ref{D: xz1 and xz2}; Proposition \ref{P: Two Line Optimal for Step 1 when x is between xd and xz1; xz2 and xu} follows from Lemma \ref{L: Two-Line and their returns above or below z}; so their proofs will be skipped.

\hfil\break {\sc Proof of Lemma \ref{L: z decreases for Three-Line Configuration}.}  Choose $-\infty<b_{1}<b_{2}\le \bar{b}=\bar{a}\le a_{2}<a_{1}<\infty$.  Let configuration $X_{1}=x_{d}\I_{A_{1}}+x\I_{B_{1}}+x_{u}\I_{D_{1}}$ correspond to the pair $(a_{1}, b_{1})$  where $A_{1}=\left\{\omega\in\Omega\,:\,\tfrac{d\tilde{P}}{dP}(\omega)>a_{1}\right\}, B_{1}=\left\{\omega\in\Omega\,:\,b_{1}\le\tfrac{d\tilde{P}}{dP}(\omega)\le a_{1}\right\}, D_{1}=\left\{\omega\in\Omega\,:\, \tfrac{d\tilde{P}}{dP}(\omega)<b_{1}\right\}$. Similarly, let configuration $X_{2}=x_{d}\I_{A_{2}}+x\I_{B_{2}}+x_{u}\I_{D_{2}}$ correspond to the pair $(a_{2}, b_{2})$.   Define $z_{1}=E[X_{1}]$ and $z_{2}=E[X_{2}]$.  Since both $X_{1}$ and $X_{2}$ satisfy the capital constraint, we have
\[x_{d}\tilde{P}(A_{1})+x\tilde{P}(B_{1})+x_{u}\tilde{P}(D_{1})=x_{r}=x_{d}\tilde{P}(A_{2})+x\tilde{P}(B_{2})+x_{u}\tilde{P}(D_{2}).\]
This simplifies to the equation
\begin{equation}\label{E: for a2}
(x-x_{d})\tilde{P}(A_{2}\backslash A_{1})=(x_{u}-x)\tilde{P}(D_{2}\backslash D_{1}).
\end{equation}
Then
\begin{align*}
z_{2}-z_{1} &=x_{d}P(A_{2})+xP(B_{2})+x_{u}P(D_{2})-x_{d}P(A_{1})-xP(B_{1})-x_{u}P(D_{1})\\
&=(x_{u}-x)P(D_{2}\backslash D_{1})-(x-x_{d})P(A_{2}\backslash A_{1})\\
&=(x_{u}-x)P(D_{2}\backslash D_{1})-(x_{u}-x)\frac{\tilde{P}(D_{2}\backslash D_{1})}{\tilde{P}(A_{2}\backslash A_{1})}P(A_{2}\backslash A_{1})\\
&=(x_{u}-x)\tilde{P}(D_{2}\backslash D_{1})\left(\frac{P(D_{2}\backslash D_{1})}{\tilde{P}(D_{2}\backslash D_{1})}
-\frac{P(A_{2}\backslash A_{1})}{\tilde{P}(A_{2}\backslash A_{1})}\right)\\
&=(x_{u}-x)\tilde{P}(D_{2}\backslash D_{1})\left(
\frac{\int_{\left\{b_{1}\le\tfrac{d\tilde{P}}{dP}(\omega)< b_{2}\right\}}\tfrac{dP}{d\tilde{P}}(\omega)d\tilde{P}(\omega)}{\tilde{P}(D_{2}\backslash D_{1})}
-\frac{\int_{\left\{a_{2}<\tfrac{d\tilde{P}}{dP}(\omega)\le a_{1}\right\}}\tfrac{dP}{d\tilde{P}}(\omega)d\tilde{P}(\omega)}{\tilde{P}(A_{2}\backslash A_{1})}\right)\\
&\ge (x_{u}-x)\tilde{P}(D_{2}\backslash D_{1})\left(\frac{1}{b_{2}}-\frac{1}{a_{2}}\right)>0.
\end{align*}
Suppose the pair $(a_{1}, b_{1})$ is chosen so that $X_{1}$ satisfies the budget constraint $\tilde{E}[X_{1}]=x_{r}$.  For any given $\epsilon>0$, choose $b_{2}-b_{1}$ small enough such that $P(D_{2}\backslash D_{1})\le \frac{\epsilon}{x_{u}-x}$.  Now choose $a_{2}$ such that $a_{2}<a_{1}$ and equation (\ref{E: for a2}) is satisfied.  Then $X_{2}$ also satisfies the budget constraint $\tilde{E}[X_{2}]=x_{r}$, and
\[z_{2}-z_{1}=(x_{u}-x)P(D_{2}\backslash D_{1})-(x-x_{d})P(A_{2}\backslash A_{1})\le (x_{u}-x)P(D_{2}\backslash D_{1})\le\epsilon.\]
We conclude that the expected value of the Three-Line configuration decreases continuously as $b$ decreases and $a$ increases.
$\endproof$

\hfil\break In the following we provide the main proof of the paper: the optimality of the Three-Line configuration.

\hfil\break {\sc Proof of Proposition \ref{P: Three Line Optimal for Step 1 when x is between xz1 and xz2}.}  Denote $\rho=\frac{d\tilde{P}}{dP}$.  According to Lemma \ref{L: z decreases for Three-Line Configuration}, there exists a Three-Line configuration $\hat{X}=x_{d}\I_{A}+x\I_{B}+x_{u}\I_{D}$ that satisfies the General Constraints:
\begin{align*}
&E[X]=x_{d}P(A)+xP(B)+x_{u}P(D)=z, \\
&\tilde{E}[X]=x_{d}\tilde{P}(A)+x\tilde{P}(B)+x_{u}\tilde{P}(D)=x_{r}.
\end{align*}
where
\begin{equation*}
A=\left\{\omega\in\Omega\,:\,\rho(\omega)>\hat{a}\right\},\quad
B=\left\{\omega\in\Omega\,:\,\hat{b}\le\rho(\omega)\le \hat{a}\right\},\quad
D=\left\{\omega\in\Omega\,:\, \rho(\omega)<\hat{b}\right\}.
\end{equation*}
As standard for convex optimization problems, if we can find a pair of Lagrange multipliers $\lambda\ge0$ and $\mu\in\R$ such that $\hat{X}$ is the solution to the minimization problem
\begin{equation}\label{E: Langrange Multiplier}
\inf_{X\in\sF, \,\,x_{d}\le X\le x_{u}}E[(x-X)^{+}-\lambda X-\mu \rho X]=E[(x-\hat{X})^{+}-\lambda \hat{X}-\mu \rho \hat{X}],
\end{equation}
then $\hat{X}$ is the solution to the constrained problem
\[\inf_{X\in\sF, \,\,x_{d}\le X\le x_{u}}E[(x-X)^{+}], \quad s.t.\quad E[X]\ge z, \quad\tilde{E}[X]=x_{r}.\]
Define \[\lambda=\frac{\hat{b}}{\hat{a}-\hat{b}}, \quad \mu=-\frac{1}{\hat{a}-\hat{b}}.\]
Then (\ref{E: Langrange Multiplier}) becomes
\[\inf_{X\in\sF, \,\,x_{d}\le X\le x_{u}}E\left[(x-X)^{+}+\tfrac{\rho-\hat{b}}{\hat{a}-\hat{b}} X\right].
\]
Choose any $X\in\sF$ where $x_{d}\le X\le x_{u}$, and denote $G=\{\omega\in\Omega\,:\, X(\omega)\ge x\}$ and $L=\{\omega\in\Omega\,:\, X(\omega)< x\}$.  Note that $\tfrac{\rho-\hat{b}}{\hat{a}-\hat{b}}>1$ on set $A$, $0\le\tfrac{\rho-\hat{b}}{\hat{a}-\hat{b}}\le1$ on set $B$, $\tfrac{\rho-\hat{b}}{\hat{a}-\hat{b}}<0$ on set $D$. Then the difference
\begin{align*}
&\,\,E\left[(x-X)^{+}+\tfrac{\rho-\hat{b}}{\hat{a}-\hat{b}} X\right]-E\left[(x-\hat{X})^{+}+\tfrac{\rho-\hat{b}}{\hat{a}-\hat{b}} \hat{X}\right]\\
&= E\left[(x-X)\I_{L}+\tfrac{\rho-\hat{b}}{\hat{a}-\hat{b}} X\left(\I_{A}+\I_{B}+\I_{D}\right)\right]
-E\left[\left(x-x_{d}\right)\I_{A}+\tfrac{\rho-\hat{b}}{\hat{a}-\hat{b}} (x_{d}\I_{A}+x\I_{B}+x_{u}\I_{D})\right]\\
&=E\left[(x-X)\I_{L}+\left(\tfrac{\rho-\hat{b}}{\hat{a}-\hat{b}} (X-x_{d})-(x-x_{d})\right)\I_{A}
+\tfrac{\rho-\hat{b}}{\hat{a}-\hat{b}} \left(X-x\right)\I_{B}+\tfrac{\rho-\hat{b}}{\hat{a}-\hat{b}} \left(X-x_{u}\right)\I_{D}\right]\\
&\ge E\left[(x-X)\I_{L}+\left(X-x\right)\I_{A}
+\tfrac{\rho-\hat{b}}{\hat{a}-\hat{b}} \left(X-x\right)\I_{B}+\tfrac{\rho-\hat{b}}{\hat{a}-\hat{b}} \left(X-x_{u}\right)\I_{D}\right] \\
&= E\left[(x-X)\left(\I_{L\cap A}+\I_{L\cap B}+\I_{L\cap D}\right)+\left(X-x\right)\left(\I_{A\cap G}+\I_{A\cap L}\right)
+\tfrac{\rho-\hat{b}}{\hat{a}-\hat{b}} \left(X-x\right)\I_{B}+\tfrac{\rho-\hat{b}}{\hat{a}-\hat{b}} \left(X-x_{u}\right)\I_{D}\right]\\
&= E\left[(x-X)\left(\I_{L\cap B}+\I_{L\cap D}\right)+\left(X-x\right)\I_{A\cap G}
+\tfrac{\rho-\hat{b}}{\hat{a}-\hat{b}} \left(X-x\right)\I_{B}+\tfrac{\rho-\hat{b}}{\hat{a}-\hat{b}} \left(X-x_{u}\right)\I_{D}\right]\\
&= E\left[(x-X)\left(\I_{L\cap B}+\I_{L\cap D}\right)+\left(X-x\right)\I_{A\cap G}
+\tfrac{\rho-\hat{b}}{\hat{a}-\hat{b}} \left(X-x\right)\left(\I_{B\cap G}+\I_{B\cap L}\right)+\tfrac{\rho-\hat{b}}{\hat{a}-\hat{b}} \left(X-x_{u}\right)\left(\I_{D\cap G}+\I_{D\cap L}\right)\right]\\
&= E\left[(x-X)\left(1-\tfrac{\rho-\hat{b}}{\hat{a}-\hat{b}}\right)\I_{B\cap L}+\left(x-X+\tfrac{\rho-\hat{b}}{\hat{a}-\hat{b}} \left(X-x_{u}\right)\right)\I_{D\cap L}+\left(X-x\right)\I_{A\cap G}\right.\\
&\left.\qquad\qquad+\tfrac{\rho-\hat{b}}{\hat{a}-\hat{b}} \left(X-x\right)\I_{B\cap G}+\tfrac{\rho-\hat{b}}{\hat{a}-\hat{b}} \left(X-x_{u}\right)\I_{D\cap G}\right]\ge0.
\end{align*}
The last inequality holds because each term inside the expectation is greater than or equal to zero.
$\endproof$

\hfil\break Theorem \ref{T: Solution to Step 1} is a direct consequence of Lemma \ref{L: Two-Line and their returns above or below z}, Proposition \ref{P: Two Line Optimal for Step 1 when x is between xd and xz1; xz2 and xu}, and Proposition \ref{P: Three Line Optimal for Step 1 when x is between xz1 and xz2}.

\hfil\break {\sc Proof of Lemma \ref{L: convexity of v(x)}.} The convexity of $v(x)$ is a simple consequence of its definition (\ref{E: Step 1}).  Real-valued convex functions on $\R$ are continuous on its interior of the domain, so $v(x)$ is continuous on $\R$.
$\endproof$

\hfil\break {\sc Proof of Proposition \ref{P: Three-Line Optimal for Step 2 in Two-Constraint Case}.} 
For $z\in(z^{*},\zbar]$, \textbf{Step 2} of the \textbf{Two-Constraint Problem}
\[\frac{1}{\lambda}\inf_{x\in\R}(v(x)-\lambda x)\] is the minimum of the following five sub-problems after applying Theorem \ref{T: Solution to Step 1}:
\begin{description}
\item[Case 1] \[\frac{1}{\lambda} \inf_{(-\infty, x_{d}]}(v(x)-\lambda x) = \frac{1}{\lambda} \inf_{(-\infty, x_{d}]}(-\lambda x)=-x_{d}; \]
\item[Case 2] \[\frac{1}{\lambda} \inf_{[x_{d}, x_{z1}]}(v(x)-\lambda x) = \frac{1}{\lambda} \inf_{[x_{d}, x_{z1}]}(-\lambda x)=-x_{z1}\le -x_{d}; \]
\item[Case 3] \[\frac{1}{\lambda} \inf_{(x_{z1}, x_{z2})}(v(x)-\lambda x) = \frac{1}{\lambda} \inf_{(x_{z1}, x_{z2})}\left((x-x_{d})P(A_{x})-\lambda x\right);\]
\item[Case 4] \[\frac{1}{\lambda} \inf_{[x_{z2}, x_{u}]}(v(x)-\lambda x) = \frac{1}{\lambda} \inf_{[x_{z2}, x_{u}]}\left((x-x_{d})P(A_{x})-\lambda x\right);\]
\item[Case 5] \[\frac{1}{\lambda} \inf_{[x_{u}, \infty)}(v(x)-\lambda x) = \frac{1}{\lambda} \inf_{[x_{u}, \infty)}\left((x-x_{d})P(\bar{A})+(x-x_{u})P(\bar{B})-\lambda x\right).\]
\end{description}
Obviously, \textbf{Case 2} dominates \textbf{Case 1} in the sense that its minimum is lower.  In \textbf{Case 3}, by the continuity of $v(x)$, we have
\[\frac{1}{\lambda} \inf_{(x_{z1}, x_{z2})}\left((x-x_{d})P(A_{x})-\lambda x\right) \le
 \frac{1}{\lambda} \left((x_{z1}-x_{d})P(A_{x_{z1}})-\lambda x_{z1}\right)=-x_{z1}.\]
The last equality comes from the fact $P(A_{x_{z1}})=0$: As in Lemma \ref{L: z decreases for Three-Line Configuration}, we know that when $x=x_{z1}$, the Three-Line configuration $X=x_{d}\I_{A}+x\I_{B}+x_{u}\I_{D}$ degenerates to the Two-Line configuration $X=x_{z1}\I_{B}+x_{u}\I_{D}$ where $a_{x_{z1}}=\infty$.   Therefore, \textbf{Case 3} dominates \textbf{Case 2}.    In \textbf{Case 5},
\begin{align*}
\frac{1}{\lambda} \inf_{[x_{u}, \infty)}(v(x)-\lambda x) &= \frac{1}{\lambda} \inf_{[x_{u}, \infty)}\left((x-x_{d})P(\bar{A})+(x-x_{u})P(\bar{B})-\lambda x\right)\\
&=  \frac{1}{\lambda} \inf_{[x_{u}, \infty)}\left((1-\lambda)x-x_{d}P(\bar{A})-x_{u}P(\bar{B})\right)\\
&=  \frac{1}{\lambda}\left((1-\lambda)x_{u}-x_{d}P(\bar{A})-x_{u}P(\bar{B})\right)\\
&=  \frac{1}{\lambda}\left((x_{u}-x_{d})P(\bar{A})-\lambda x_{u}\right)\\
&\ge \frac{1}{\lambda} \inf_{[x_{z2}, x_{u}]}\left((x-x_{d})P(A_{x})-\lambda x\right).
\end{align*}
Therefore, \textbf{Case 4} dominates \textbf{Case 5}.    When $x\in[x_{z2}, x_{u}]$ and $\esssup\frac{d\tilde{P}}{dP}>\frac{1}{\lambda}$, Theorem \ref{T: Solution to Step 1} and Theorem \ref{T: Two Line Optimal in One Constraint Sase} imply that the infimum in \textbf{Case 4} is achieved either by $\bar{X}$ or $X^{*}$.  Since we restrict $z\in(z^{*}, \zbar]$ where $z^{*}=\zbar$ by Definition \ref{D: z*} in the first case, we need not consider this case in the current proposition. In the second case, Lemma \ref{L: z decreases when x is between xd and xr and increases when x is between xr and xu} implies that $x^{*}< x_{z2}$ (because $z>z^{*}$).  By the convexity of $v(x)$, and then the continuity of $v(x)$,
\begin{align*}
\frac{1}{\lambda} \inf_{[x_{z2}, x_{u}]}\left((x-x_{d})P(A_{x})-\lambda x\right) &=\frac{1}{\lambda} \left((x_{z2}-x_{d})P(A_{x_{z2}})-\lambda x_{z2}\right)\\
&\ge \frac{1}{\lambda} \inf_{(x_{z1}, x_{z2})}\left((x-x_{d})P(A_{x})-\lambda x\right).
\end{align*}
Therefore, \textbf{Case 3} dominates \textbf{Case 4}.  We have shown that \textbf{Case 3} actually provides the globally infimum:
\[\frac{1}{\lambda}\inf_{x\in\R}(v(x)-\lambda x)=\frac{1}{\lambda} \inf_{(x_{z1}, x_{z2})}(v(x)-\lambda x).\]
Now we focus on $x\in(x_{z1}, x_{z2})$, where $X(x)=x_{d}\I_{A_{x}}+x\I_{B_{x}}+x_{u}\I_{D_{x}}$ satisfies the general constraints:
\begin{align*}
&E[X(x)]=x_{d}P(A_{x})+xP(B_{x})+x_{u}P(D_{x})=z, \\
&\tilde{E}[X(x)]=x_{d}\tilde{P}(A_{x})+x\tilde{P}(B_{x})+x_{u}\tilde{P}(D_{x})=x_{r},
\end{align*}
and the definition for sets $A_{x}$, $B_{x}$ and $D_{x}$ are
\begin{equation*}
A_{x}=\left\{\omega\in\Omega\,:\,\tfrac{d\tilde{P}}{dP}(\omega)>a_{x}\right\},\quad
B_{x}=\left\{\omega\in\Omega\,:\,b_{x}\le\tfrac{d\tilde{P}}{dP}(\omega)\le a_{x}\right\},\quad
D_{x}=\left\{\omega\in\Omega\,:\, \tfrac{d\tilde{P}}{dP}(\omega)<b_{x}\right\}.
\end{equation*}
Note that $v(x)=(x-x_{d})P(A_{x})$ (see Theorem \ref{T: Solution to Step 1}).
Since $P(A_{x})+P(B_{x})+P(D_{x})=1$ and $\tilde{P}(A_{x})+\tilde{P}(B_{x})+\tilde{P}(D_{x})=1$, we rewrite the capital and return constraints as
\begin{align*}
x-z &=(x-x_{d})P(A_{x})+(x-x_{u})P(D_{x}), \\
x-x_{r} &=(x-x_{d})\tilde{P}(A_{x})+(x-x_{u})\tilde{P}(D_{x}).
\end{align*}
Differentiating both sides with respect to $x$, we get
\begin{align*}
P(B_{x}) &=(x-x_{d})\frac{dP(A_{x})}{dx}+(x-x_{u})\frac{dP(D_{x})}{dx}, \\
\tilde{P}(B_{x}) &=(x-x_{d})\frac{d\tilde{P}(A_{x})}{dx}+(x-x_{u})\frac{d\tilde{P}(D_{x})}{dx}.
\end{align*}
Since
\[\frac{d\tilde{P}(A_{x})}{dx}=a_{x}\frac{dP(A_{x})}{dx}, \quad \frac{d\tilde{P}(D_{x})}{dx}=b_{x}\frac{dP(D_{x})}{dx},\]
we get
\[\frac{dP(A_{x})}{dx}=\frac{\tilde{P}(B_{x})-bP(B_{x})}{(x-x_{d})(a-b)}.\]
Therefore,
\begin{align*}
(v(x)-\lambda x)' &=P(A_{x})+(x-x_{d})\frac{dP(A_{x})}{dx}-\lambda\\
&=P(A_{x})+\frac{\tilde{P}(B_{x})-bP(B_{x})}{a-b}-\lambda.
\end{align*}
When the above derivative is zero, we arrive to the first order Euler condition
\[P(A_{x})+\frac{\tilde{P}(B_{x})-bP(B_{x})}{a-b}-\lambda=0.\]
To be precise, the above differentiation should be replaced by left-hand and right-hand derivatives as detailed in the Proof for Corollary 2.8 in Li and Xu \cite{LiXuA}.  But the first order Euler condition will turn out to be the same because we have assumed that the Radon-Nikod\'{y}m derivative $\frac{d\tilde{P}}{dP}$ has continuous distribution.

To finish this proof, we need to show that there exists an $x\in (x_{z1}, x_{z2})$ where the first order Euler condition is satisfied.  From Lemma \ref{L: z decreases for Three-Line Configuration}, we know that as $x\searrow x_{z1}$, $a_{x}\nearrow\infty$, and $P(A_{x})\searrow0$.  Therefore,
\[\lim_{x\searrow x_{z1}}(v(x)-\lambda x)' = -\lambda <0.\]
As $x\nearrow x_{z2}$, $b_{x}\searrow0$, and $P(D_{x})\searrow0$.  Therefore,
\[\lim_{x\nearrow x_{z2}}(v(x)-\lambda x)' = P(A_{x_{z2}})-\frac{\tilde{P}(A_{x_{z2}}^{c})}{a_{x_{z2}}}-\lambda.\]
This derivative coincides with the derivative of the value function of the Two-Line configuration that is
optimal on the interval $x\in[x_{z2}, x_{u}]$ provided in Theorem \ref{T: Solution to Step 1} (see Proof for
Corollary 2.8 in Li and Xu \cite{LiXuA}). Again when $x\in[x_{z2}, x_{u}]$ and
$\esssup\frac{d\tilde{P}}{dP}>\frac{1}{\lambda}$, Theorem \ref{T: Solution to Step 1} and Theorem \ref{T: Two
Line Optimal in One Constraint Sase} imply that the infimum of $v(x)-\lambda x$ is achieved either by $\bar{X}$
or $X^{*}$.  Since we restrict $z\in(z^{*}, \zbar]$ where $z^{*}=\zbar$ by Definition \ref{D: z*} in the first
case, we need not consider this case in the current proposition. In the second case, Lemma \ref{L: z decreases
when x is between xd and xr and increases when x is between xr and xu} implies that $x^{*}< x_{z2}$ (because
$z>z^{*}$).  This in turn implies
\[P(A_{x_{z2}})-\frac{\tilde{P}(A_{x_{z2}}^{c})}{a_{x_{z2}}}-\lambda<0.\]
We have just shown that there exist some $x^{**}\in(x_{z1}, x_{z2})$ such that $(v(x)-\lambda x)'|_{x=x^{**}}=0$.  By the convexity of $v(x)-\lambda x$, this is the point where it obtains the minimum value.  Now
\begin{align*}
CVaR (X^{**}) &=\frac{1}{\lambda}\left(v(x^{**})-\lambda x^{**}\right)\\
&= \frac{1}{\lambda}\left((x^{**}-x_{d})P(A^{**})-\lambda x^{**}\right).
\end{align*}
$\endproof$

\hfil\break {\sc Proof of Theorem \ref{T: Solution to Step 2}.}
Case 3 and 4 are already proved in Theorem \ref{T: Two Line Optimal in One Constraint Sase} and Proposition \ref{P: Three-Line Optimal for Step 2 in Two-Constraint Case}.  In Case 1 where $\esssup\frac{d\tilde{P}}{dP}\le\frac{1}{\lambda}$ and $z=x_{r}$,  $X=x_{r}$  is both feasible and optimal by Theorem \ref{T: Two Line Optimal in One Constraint Sase}.  In Case 2, fix arbitrary $\epsilon>0$. We will look for a Two-Line solution $X_{\epsilon}=x_{\epsilon}\I_{A_{\epsilon}}+\alpha_{\epsilon}\I_{B_{\epsilon}}$ with the right parameters $a_{\epsilon}, x_{\epsilon}, \alpha_{\epsilon}$ which satisfies both the capital constraint and return constraint:
\begin{align}
&E[X_{\epsilon}]=x_{\epsilon}P(A_{\epsilon})+\alpha_{\epsilon} P(B_{\epsilon})=z, \label{E: z for epsilon two line}\\
&\tilde{E}[X_{\epsilon}]=x_{\epsilon}\tilde{P}(A_{\epsilon})+\alpha_{\epsilon}\tilde{P}(B_{\epsilon})=x_{r}, \label{E: xr for epsilon two line}
\end{align}
where
\begin{equation*}
A_{\epsilon}=\left\{\omega\in\Omega\,:\,\tfrac{d\tilde{P}}{dP}(\omega)>a_{\epsilon}\right\},\quad
B_{\epsilon}=\left\{\omega\in\Omega\,:\,\tfrac{d\tilde{P}}{dP}(\omega)\le a_{\epsilon}\right\},
\end{equation*}
and produces a CVaR level close to the lower bound:
\[CVaR(X_{\epsilon})\le CVaR(x_{r})+\epsilon=-x_{r}+\epsilon.\]
First, we choose $x_{\epsilon}=x_{r}-\epsilon$.  To find the remaining two parameters $a_{\epsilon}$ and $\alpha_{\epsilon}$ so that equations (\ref{E: z for epsilon two line}) and (\ref{E: xr for epsilon two line}) are satisfies, we note
\begin{align*}
&x_{r}P(A_{\epsilon})+x_{r} P(B_{\epsilon})=x_{r},\\
&x_{r}\tilde{P}(A_{\epsilon})+x_{r}\tilde{P}(B_{\epsilon})=x_{r},
\end{align*}
and conclude that it is equivalent to find a pair of $a_{\epsilon}$ and $\alpha_{\epsilon}$ such that the following two equalities are satisfied:
\begin{align*}
-\epsilon P(A_{\epsilon})+(\alpha_{\epsilon}-x_{r}) P(B_{\epsilon})&=\gamma,\\
-\epsilon \tilde{P}(A_{\epsilon})+(\alpha_{\epsilon}-x_{r})\tilde{P}(B_{\epsilon}) &=0,
\end{align*}
where we denote $\gamma=z-x_{r}$.  If we can find a solution $a_{\epsilon}$ to the equation
\begin{equation}\label{E: b_epsilon two line}
\frac{\tilde{P}(B_{\epsilon})}{P(B_{\epsilon})}=\frac{\epsilon}{\gamma+\epsilon},
\end{equation}
then \[\alpha_{\epsilon}=x_{r}+\frac{\tilde{P}(A_{\epsilon})}{\tilde{P}(B_{\epsilon})}\epsilon,\] and we have the solutions for equations (\ref{E: z for epsilon two line}) and (\ref{E: xr for epsilon two line}).  It is not difficult to prove that the fraction $\frac{\tilde{P}(B)}{P(B)}$ increases continuously from $0$ to $1$ as $a$ increases from $0$ to $\frac{1}{\lambda}$.  Therefore, we can find a solution $a_{\epsilon}\in(0, \frac{1}{\lambda})$ where (\ref{E: b_epsilon two line}) is satisfied.  By definition (\ref{D: CVaR}),
\[
CVaR_{\lambda}(X_{\epsilon}) =\frac{1}{\lambda}\inf_{x\in\R}\left(E[(x-X_{\epsilon})^{+}]-\lambda x\right)
\le \frac{1}{\lambda}\left(E[(x_{\epsilon}-X_{\epsilon})^{+}]-\lambda x_{\epsilon}\right)=-x_{\epsilon}.
\]
The difference
\[
CVaR_{\lambda}(X_{\epsilon})-CVaR(x_{r}) \le -x_{\epsilon}+x_{r}=\epsilon.
\]
Under Assumption \ref{A: Complete Market and Continuous RND}, the solution in Case 2 is almost surely unique,  the result is proved.
$\endproof$

\hfil\break {\sc Proof of Theorem \ref{T: unbounded above}.}
Case 1 and 3 are obviously true in light of Theorem \ref{T: Two Line Optimal in One Constraint Case Infinite Upper Bound}.  The proof for Case 2 is similar to that in the Proof of Theorem \ref{T: Solution to Step 2}, so we will not repeat it here.  Since $E[X^{*}]=z^{*}<z$ in case 4, $CVaR(X^{*})$ is only a lower bound in this case.  We first show that it is the true infimum obtained in Case 4. Fix arbitrary $\epsilon>0$. We will look for a Three-Line solution $X_{\epsilon}=x_{d}\I_{A_{\epsilon}}+x_{\epsilon}\I_{B_{\epsilon}}+\alpha_{\epsilon}\I_{D_{\epsilon}}$ with the right parameters $a_{\epsilon}, b_{\epsilon}, x_{\epsilon}, \alpha_{\epsilon}$ which satisfies the general constraints:
\begin{align}
&E[X_{\epsilon}]=x_{d}P(A_{\epsilon})+x_{\epsilon}P(B_{\epsilon})+\alpha_{\epsilon} P(D_{\epsilon})=z, \label{E: z for epsilon}\\
&\tilde{E}[X_{\epsilon}]=x_{d}\tilde{P}(A_{\epsilon})+x_{\epsilon}\tilde{P}(B_{\epsilon})+\alpha_{\epsilon}\tilde{P}(D_{\epsilon})=x_{r}, \label{E: xr for epsilon}
\end{align}
where
\begin{equation*}
A_{\epsilon}=\left\{\omega\in\Omega\,:\,\tfrac{d\tilde{P}}{dP}(\omega)>a_{\epsilon}\right\},\quad
B_{\epsilon}=\left\{\omega\in\Omega\,:\,b_{\epsilon}\le\tfrac{d\tilde{P}}{dP}(\omega)\le a_{\epsilon}\right\},\quad
D_{\epsilon}=\left\{\omega\in\Omega\,:\, \tfrac{d\tilde{P}}{dP}(\omega)<b_{\epsilon}\right\},
\end{equation*}
and produces a CVaR level close to the lower bound:
\[CVaR(X_{\epsilon})\le CVaR(X^{*})+\epsilon.\]
First, we choose $a_{\epsilon}=a^{*}$, $A_{\epsilon}=A^{*}$, $x_{\epsilon}=x^{*}-\delta$, where we define $\delta=\frac{\lambda}{\lambda-P(A^{*})}\epsilon$.  To find the remaining two parameters $b_{\epsilon}$ and $\alpha_{\epsilon}$ so that equations (\ref{E: z for epsilon}) and (\ref{E: xr for epsilon}) are satisfies, we note
\begin{align*}
&E[X^{*}] =x_{d}P(A^{*})+x^{*}P(B^{*})=z^{*},\\
&\tilde{E}[X^{*}]=x_{d}\tilde{P}(A^{*})+x^{*}\tilde{P}(B^{*})=x_{r},
\end{align*}
and conclude that it is equivalent to find a pair of $b_{\epsilon}$ and $\alpha_{\epsilon}$ such that the following two equalities are satisfied:
\begin{align*}
-\delta (P(B^{*})-P(D_{\epsilon}))+(\alpha_{\epsilon}-x^{*}) P(D_{\epsilon})&=\gamma,\\
-\delta(\tilde{P}(B^{*})-\tilde{P}(D_{\epsilon}))+(\alpha_{\epsilon}-x^{*})\tilde{P}(D_{\epsilon}) &=0,
\end{align*}
where we denote $\gamma=z-z^{*}$.  If we can find a solution $b_{\epsilon}$ to the equation
\begin{equation}\label{E: b_epsilon}
\frac{\tilde{P}(D_{\epsilon})}{P(D_{\epsilon})}=\frac{\tilde{P}(B^{*})}{\frac{\gamma}{\delta}+P(B^{*})},
\end{equation}
then \[\alpha_{\epsilon}=x^{*}+\left(\frac{\tilde{P}(B^{*})}{\tilde{P}(D_{\epsilon})}-1\right)\delta,\] and we have the solutions for equations (\ref{E: z for epsilon}) and (\ref{E: xr for epsilon}).  It is not difficult to prove that the fraction $\frac{\tilde{P}(D)}{P(D)}$ increases continuously from $0$ to $\frac{\tilde{P}(B^{*})}{P(B^{*})}$ as $b$ increases from $0$ to $a^{*}$.  Therefore, we can find a solution $b_{\epsilon}\in(0, a^{*})$ where (\ref{E: b_epsilon}) is satisfied.
By definition (\ref{D: CVaR}),
\begin{align*}
CVaR_{\lambda}(X_{\epsilon}) &=\frac{1}{\lambda}\inf_{x\in\R}\left(E[(x-X_{\epsilon})^{+}]-\lambda x\right)\\
&\le \frac{1}{\lambda}\left(E[(x_{\epsilon}-X_{\epsilon})^{+}]-\lambda x_{\epsilon}\right)\\
&=\frac{1}{\lambda}(x_{\epsilon}-x_{d})P(A_{\epsilon})-x_{\epsilon}.
\end{align*}
The difference
\begin{align*}
CVaR_{\lambda}(X_{\epsilon})-CVaR(X^{*}) &\le \frac{1}{\lambda}(x_{\epsilon}-x_{d})P(A_{\epsilon})-x_{\epsilon}-\frac{1}{\lambda}(x^{*}-x_{d})P(A^{*})+x^{*}\\
&=\frac{1}{\lambda}(x^{*}-x_{d})(P(A_{\epsilon})-P(A^{*}))+\left(1-\frac{P(A_{\epsilon})}{\lambda}\right)(x^{*}-x_{\epsilon})=\epsilon.
\end{align*}
Under Assumption \ref{A: Complete Market and Continuous RND}, the solution in Case 4 is almost surely unique,  the result is proved.
$\endproof$



\end{document}